\documentclass[reprint,
 amsmath,amssymb,
aps,superscriptaddress,longbibliography,twocolumn,floatfix
 ]{revtex4-2}
\usepackage{graphicx}
\usepackage{dcolumn}
\usepackage{mathtools}
\usepackage{mathrsfs}

\usepackage{bm}
\usepackage{color}

\renewcommand{\figurename}{{\bf Fig.}}
\setcitestyle{open={(},close={)}}
\renewcommand{\tablename}{{\bf Table}}
\renewcommand{\thetable}{\arabic{table}}

\makeatletter
\def\fnum@figure{{\bf Fig. \thefigure}}
\def\fnum@table{{\bf Table \thetable}}
\makeatother

\begin{document}

\title{Computing with vortices: \\
Bridging fluid dynamics and its information-processing capability}

\author{Ken Goto}
\affiliation{Division of Mathematical and Physical Sciences,
Kanazawa University, Kakuma, Kanazawa 920-1192, Japan. \\}%

\author{Kohei Nakajima}
\altaffiliation{Corresponding author. \\
 {\quad} k\_nakajima@mech.t.u-tokyo.ac.jp\\
 }
\affiliation{Graduate School of Information Science and Technology, The University of Tokyo, Bunkyo-ku, 113-8656 Tokyo, Japan.\\}

\author{Hirofumi Notsu}
\affiliation{Faculty of Mathematics and Physics, Kanazawa University, Kakuma, Kanazawa  920-1192, Japan. \\}
\affiliation{PRESTO, Japan Science and Technology Agency, Kawaguchi, Saitama 332-0012, Japan.\\}

\begin{abstract}
Herein, the Karman vortex system is considered to be a large recurrent neural network, and the computational capability is numerically evaluated by emulating nonlinear dynamical systems and the memory capacity. Therefore, the Reynolds number dependence of the Karman vortex system computational performance is revealed and the optimal computational performance is achieved near the critical Reynolds number at the onset of Karman vortex shedding, which is associated with a Hopf bifurcation. Our finding advances the understanding of the relationship between the physical properties of fluid dynamics and its computational capability as well as provides an alternative to the widely believed viewpoint that the information processing capability becomes optimal at the edge of chaos.
\end{abstract}

\maketitle
\clearpage
\section*{INTRODUCTION}

Fluids can be universally observed in nature and exhibit rich and diverse dynamics as well as instabilities that form a source of inspiration for several mathematicians, physicists, engineers, and biologists in the field of nonlinear science. 
In this study, we focus on the diverse nature of fluid dynamics and make a novel attempt to elucidate its information processing capability where the meaning of information processing is the approximation of a function such as that found in neural networks.
\par
\begin{figure}[t]
 \begin{center}
  \includegraphics[width=89mm]{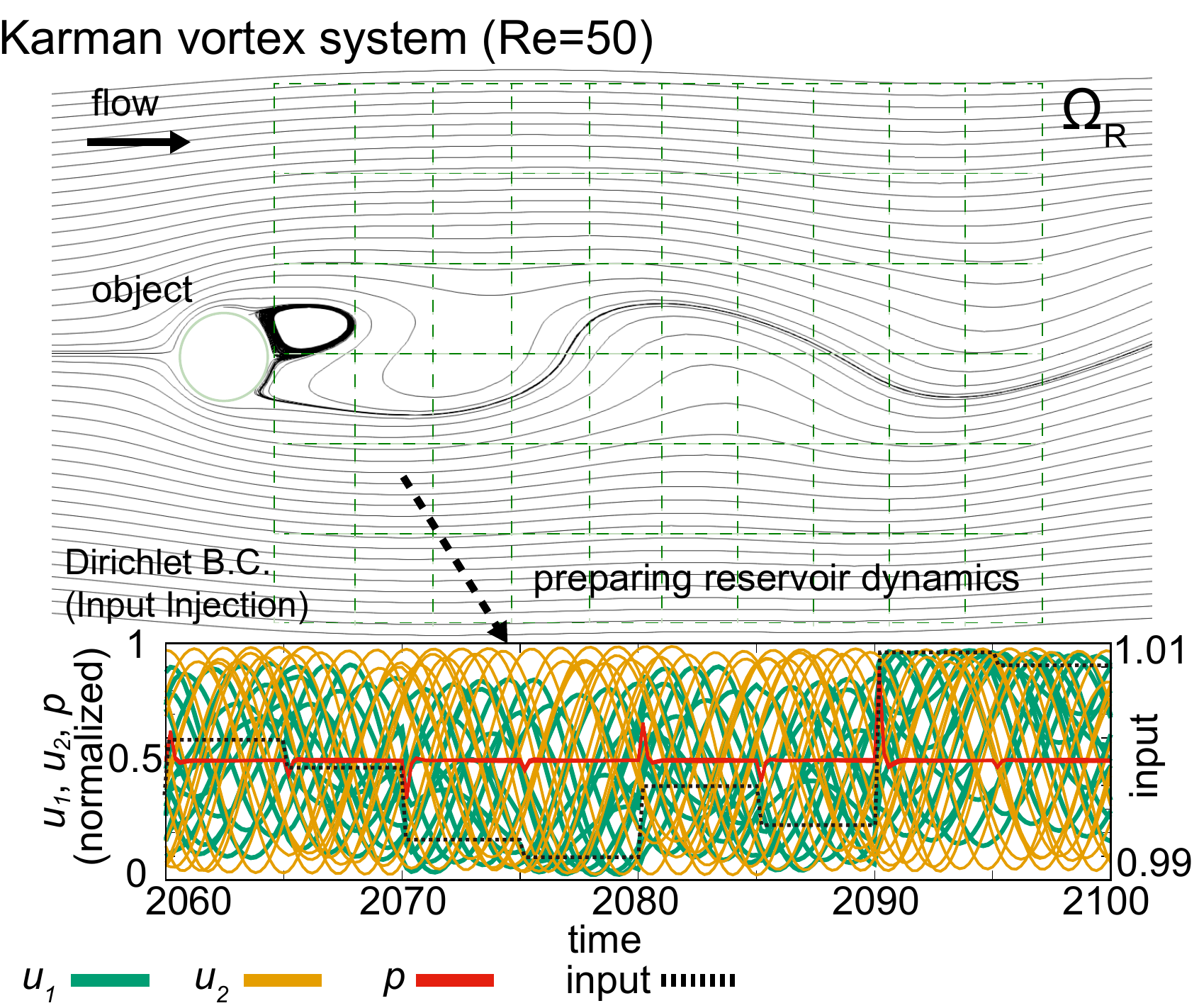}
   \end{center}
 \caption{{\bf Behavior of a typical Karman vortex, and the node preparation of reservoir for a given input sequence.} The grid behind the cylinder represents the nodes used for the reservoir; let the set of nodes be $\Omega_{R}$. The time series plots are described to be the normalized results (velocity:  ${\bf u}$ and pressure: $p$) of the fluid simulation and the input.
However, rather than using all such time series in the reservoir, the time data immediately before the switching to the next input are extracted and used. The number of output/observation nodes is $(10+1)\times(6+1)\times3=231$. All these diagrams are drawn under the conditions of $Re=50$ and $z_{1}(t)\in[0.99,1.01]$.}
\label{fig1}
\end{figure}

The Karman vortex is a renowned phenomenon in fluid dynamics and can be referred to an asymmetric vortex street that collides with an obstacle in the direction of fluid travel and is generated in the wake of an object (Fig.~\ref{fig1}), where a typical example considered to be the flow past a circular cylinder can be referred to as the Karman vortex system. This mechanism has attracted the attention of many researchers{\it\cite{Oertel,Huerre-Monk,Takemoto-Mizushima,Chomaz}}. In general, the flow at a low Reynolds number is laminar and changes to a turbulent flow as the Reynolds number increases. Moreover, this is true for the Karman vortex system, where the flow is steady and symmetric at low Reynolds numbers, and a pair of vortices, known as the twin vortex, are generated behind the object at large Reynolds numbers. The twin vortex grows in proportion with the Reynolds number; however, when the Reynolds number exceeds a certain value, vibration occurs downstream, resulting in a Karman vortex in which two rows of vortices are alternately arranged (cf. Fig.~\ref{fig2}A and supplementary videos). In this study, the Karman vortex system is used as a test bed for clarifying the computational performance of fluid dynamics.
\par
Several attempts have been made to implement computations that focus on fluids. The construction of logic circuits using the fluidity of a droplet and its application to soft robots are typical examples{\it\cite{Prakash-Ger,Katsikis-Cyb,Wehner-Tru}}. In this study, we propose another paradigm of information processing based on fluids. Here, we manage a machine learning framework called reservoir computing (RC){\it\cite{Jaeger-Haas1,maass1,Verstra-Sch}}, which is a framework of recurrent neural network learning and an information processing technique that exploits the input-driven transient behaviors of high-dimensional dynamic systems called a reservoir. This method can be implemented by adjusting linear and static readout weights from a reservoir. Moreover, an arbitrary time series can be accurately generated when the reservoir exhibits sufficient nonlinearity and memory{\it\cite{Dambre-Verstra}}. A recent development in RC is using the dynamics of a physical system as a reservoir, which can be referred to as physical reservoir computing (PRC). Examples of this implementation have been reported, including studies that have used the dynamics of water surface waves for pattern recognition{\it\cite{Fernando-Sojak}}, the usage of optical media{\it\cite{optical1,optical2}}, the behavior of a soft robot{\it\cite{Nakajima1,Nakajima2,Nakajima3,Nakajima4}}, spintronics devices{\it\cite{Torrejon-Riou,Furuta-Fujii,Tsunegi-Taniguchi}}, and the dynamics of quantum many-body systems{\it\cite{Fujii-Nakajima,Nakajima-Fujii}}, where each platform exhibits a particular computational property intrinsic to its respective spatio-temporal scales.
\par
In this study, we exploit the dynamics of the Karman vortex system as a reservoir to solve temporal machine learning tasks using numerical experiments and predict the expected outputs, that is, the fluid computational capability is analyzed by numerical simulation. 
First, we investigate the bifurcation of our Karman vortex system using input settings to understand the dynamic property of our reservoir.
Second, we analyze the echo state property (ESP), which provides a prerequisite for dynamics to function as a successful reservoir and demonstrate the relation between the ESP and the stability of the solution with respect to the Karman vortex system. 
Third, we investigate the long-diameter of twin vortices because we observe the long-diameter oscillation by the input effects to the system and analyze the behavior of the vortices satisfying the ESP. 
Next, we implement two typical benchmark tasks by applying our Karman vortex system as a reservoir to demonstrate the characteristics, the range of computational capability, the Reynolds number computational performance dependence, and its limitations. 
Finally, as an illustration of the application scenario of our scheme, we implement the prediction of fluid variables by using PRC and demonstrate its performance.
\par
This was only possible by using partial differential equations (PDE), the Navier--Stokes equations, which is crucial for our analyses to reveal vortices' detailed structure and its relation to inputs. These analyses are difficult to conduct only by using abstract spatially extended models (e.g., cell automata) in principle.

\section*{RESULTS}
In this section, our results are discussed in some subsections. 
First, we analyze the bifurcation of our system.
Next, we study ESP, which is a prerequisite to be a successful reservoir, from the perspective of a dynamical system. 
In the third subsection, we focus on the twin vortices in the region where their synchronization does not collapse according to the increase of the Reynolds number and investigate their long-diameter and the input-sensitivity inside the vortices.
In the forth and fifth subsections, the computational performances of our system for two tasks, the evaluation of the memory capacity and the nonlinear autoregressive moving average (NARMA) task, are discussed, respectively. In these tasks, we discuss the results obtained through the LR model~$y_{k}=w^1_{LR}z_{k}+w^0_{LR}$ and by a similar reservoir of flow simulation without an object (called no-object) for comparison; in this study, we refer to the results of the analysis using Navier--Stokes problem as ``system'' or ``system output''.
Finally, the results of the time series predictions for the Navier--Stokes flow, which can be applied to the interpolations of the missing variables, are denoted. This task shows whether the time series of a missing variable can be predicted by the time series of other variables without constructing their predictive models.

\subsection*{The Hopf bifurcation of the Karman vortex system}
In this subsection, we investigate the bifurcation of our system and the periodicity of the solution, where our setup is based on the non-stationary solution with inputs.
The bifurcation structure of the cylinder wake without input was first studied theoretically and numerically by Du\v{s}ek and Le Gal{\it\cite{DUS}} who revealed it to be a Hopf bifurcation. To date, however, the bifurcation structure of the system with input has not been clarified.
Fig.~\ref{fig2}A shows the typical behavior of vortices according to each Reynolds number in our system. The twin vortex ($Re=10$ and $Re=40$) and the Karman vortex ($Re=50$ and $Re=100$) are observed.  Figs.~\ref{fig2}B and C show the behavior of the circular cylinder flow using the max/min-velocity diagram and the phase portrait, respectively. We then find that the velocity is the {\it periodic type} solution at $Re=50$ and $100$, i.e., our system has non-chaotic behavior, and the Hopf bifurcation occurs near the critical Reynolds number (around $Re=45$). This result is similar to the Karman vortex system without inputs. 

\subsection*{Common-signal-induced synchronization and echo state property}
\begin{figure*}[!t]
 \begin{center}
  \includegraphics[width=183mm]{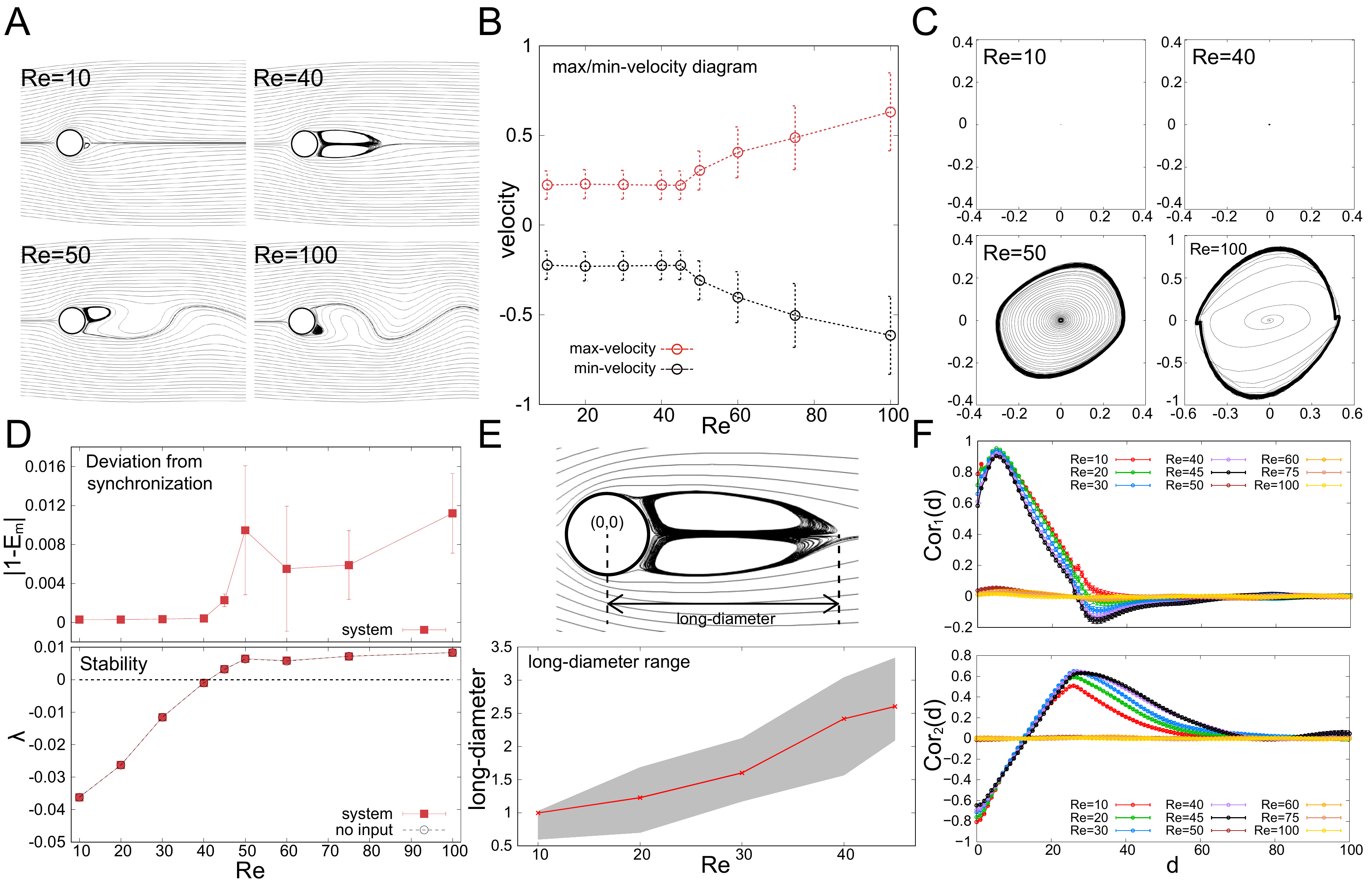}
 \end{center}
 \caption{{\bf Behaviors of fluid and the analysis of the bifurcation, and ESP, and the twin vortices' long-diameters, and their input-sensitivity, according to the Reynolds number.} ({\bf A}) The vortex types denote fluid behavior according to the Reynolds number. ({\bf B}) Denoting the Hopf bifurcation using the velocity $u_2^r$. The mean and deviation for $\max_{P\in \Omega_R, k\in[0,T_r-1]} u_{2}^r(P,k)$ and $\min_{P\in \Omega_R, k\in[0,T_r-1]} u_{2}^r(P,k)$ are plotted according to the Reynolds number. ({\bf C}) The phase portrait $(u_{2},\partial u_{2}/\partial t)$, which is constructed at $(x_1,x_2)=(5.5,0.0)$ and $Re=10,40,50,$ and $100$. ({\bf D}) The mean and deviation for $|1-E_m|$ and $\lambda$ are plotted according to the Reynolds number. 
({\bf E}) The overview of the twin vortices' long-diameter and the long-diameter range according to the Reynolds number. The red line represents the averaged long-diameter and the gray fill indicates the max/min long-diameter at $Re=10,20,30,40$, and $45$. ({\bf F}) The mean and deviation cross-correlations $Cor_i(d),i=1,2$, which between the flow function at the two fixed points $(x_1,x_2)\approx(0.52,-0.2),(3.3,-0.32)$, and the inputs are plotted according to the delay $d$  by the Reynolds number.}
 \label{fig2}
\end{figure*}
When a signal-driven dynamical system is used as a successful reservoir, the dynamical system must satisfy common-signal-induced synchronization{\it\cite{maass1,Toral-Mirasso,Lu-Hunt}} (or ESP{\it\cite{Yildiz-Jaeger,Manjunath-Jaeger}}).  
Then, we analyze two numbers, $E_m$ and $\lambda$, for each Reynolds number, where $E_m$ represents an indicator of synchronization and asymptotically converges to $1$ if synchronization occurs, and $\lambda$ is the amplification factor of perturbation and means the stability/instability of the flow field, cf. the section of MATERIALS AND METHODS for the definitions of $E_m$ and $\lambda$.
Fig.~\ref{fig2}D shows the outcome of the averaged $|1-E_m|$ and $\lambda$ in each Reynolds number.  Both $|1-E_m|$ and $\lambda$ tend to increase depending on the Reynolds number, suggesting a critical point ($\in[40,50]$) where synchronization breaks down during the transition from the twin vortex to the Karman vortex. 
The synchronization phenomenon is unlikely to occur under the condition of $Re\geq 50$; i.e., the Karman vortex is generated. 
However, in the region where the synchronization is collapsed ($Re\in (45,100]$), no chaotic behavior occurs but it is a {\it periodic} behavior (Fig.~\ref{fig2}B and Fig.~\ref{fig2}C).
Furthermore, the onset of the flow instability near the critical Reynolds number has been known to be a manifestation of Hopf bifurcation and a chaotic behavior appears at a larger $Re$ (cf., e.g.,{\it\cite{DUS,WILL}}). Therefore, note that the critical point of the synchronization differs from that of {\it chaos} of flow field.
In our setup, $z_{1}$ uses an input with a small swing, such as $[0.99,1.01]$; thus, $\lambda$ is almost the same for systems without input (Fig.~\ref{fig2}D). Interestingly, the behaviors of $|1-E_m|$ and $\lambda$ correspond to the task results presented in Fig.~\ref{fig3}, which will be explained in detail in a later section.

\subsection*{Twin vortex behavior}
The twin vortices are known to grow large according to the Reynolds number. In this subsection, we calculate the twin vortices' long-diameter using the flow function. 
The behavior of this long-diameter becomes the brief indicator of the input effects because the long-diameter is observed to be oscillating in response to the input stream (the supplementary videos). 
To clarify the vortex internal structures of our system, we focus on the twin vortices, i.e., $Re\in [10, 45]$ and analyze the input-sensitivity of the twin vortices here. Primarily, the twin vortices are described, but as a comparison, the Karman vortex cases are described too.
Let us introduce the twin vortices' long-diameter:
\begin{align*}
\gamma(t) & \coloneqq \max_{{\bf x}\in \Omega} \{x_1; \psi(t)>0, x_2<0 \},\\
{\tilde\gamma}(t) & \coloneqq \max_{{\bf x}\in \Omega} \{x_1; \psi(t)<0, x_2>0 \},
\end{align*}
where $\psi$ is the flow function satisfied $u_1=\partial \psi/\partial x_2, u_2=-\partial \psi/\partial x_1$. In Fig.~\ref{fig2}E, the averaged $(\gamma+{\tilde{\gamma}})/2$ and max/min $\gamma$ or ${\tilde\gamma}$ for time are plotted according to the Reynolds number. 
We observe the oscillation of the long-diameter by the numerical simulation, and then the long-diameter is calculated. Subsequently, we determine the expanded long-diameter range with the input effects (Fig.~\ref{fig2}E). 
Next, we calculate the cross-correlation between the inputs $z^n=z_1(n\Delta t)$ and the flow function $\psi_1^n=\psi({\bf x}_{min},n\Delta t)$,  $\psi_2^n=\psi({\bf x}_{max},n\Delta t)$ where ${\bf x}_{min}\approx(0.52,-0.2), {\bf x}_{max}\approx(3.3,-0.32)$. These coordinates represent the coordinate near the cylinder where the vortices are generated and the coordinate taking approximately the long-diameter at $Re=45$, respectively.
In Fig.~\ref{fig2}F, we use the indicator $Cor_i(d), i=1,2$ as follows:\par
\begin{align*}
Cor_i(d)  \coloneqq \frac{cov(\psi_i^{k+d},z^k)}{s(\psi_i^{k+d})s(z^k)}, i=1,2, \\
\end{align*}
where $cov(\psi_i^{k+d},z^k)$ and $s(\psi_i^{k+d})$ represent the covariance between $\psi_i^{k+d}$ and $z^k$ and the standard deviation of $\psi_i^{k+d}$, respectively, and the delay $d=0,\dots,100$. 
In the beginning, there is no large difference between the input-sensitivity at ${\bf x}_{min}$ according to the Reynolds number $Re\in[10,45]$  but the higher Reynolds number, the slower convergence of $Cor_i(d)$. 
Moreover, the input-sensitivity at ${\bf x}_{max}$ is more significantly different according to the Reynolds number $Re\in[10,45]$.
This indicates that the system has wider effective vortices with respect to the input-sensitivity in proportion with the Reynolds number $Re\in[10,45]$, which this forms an essential property of the vortices reservoir. 
In the case of the Karman vortex $Re\in[50,100]$, the input-sensitivity is exceedingly low at both points, suggesting that the ESP has collapsed.

\subsection*{Evaluation of the memory capacity}
\begin{figure*}[!t]
 \begin{center}
  \includegraphics[width=183mm]{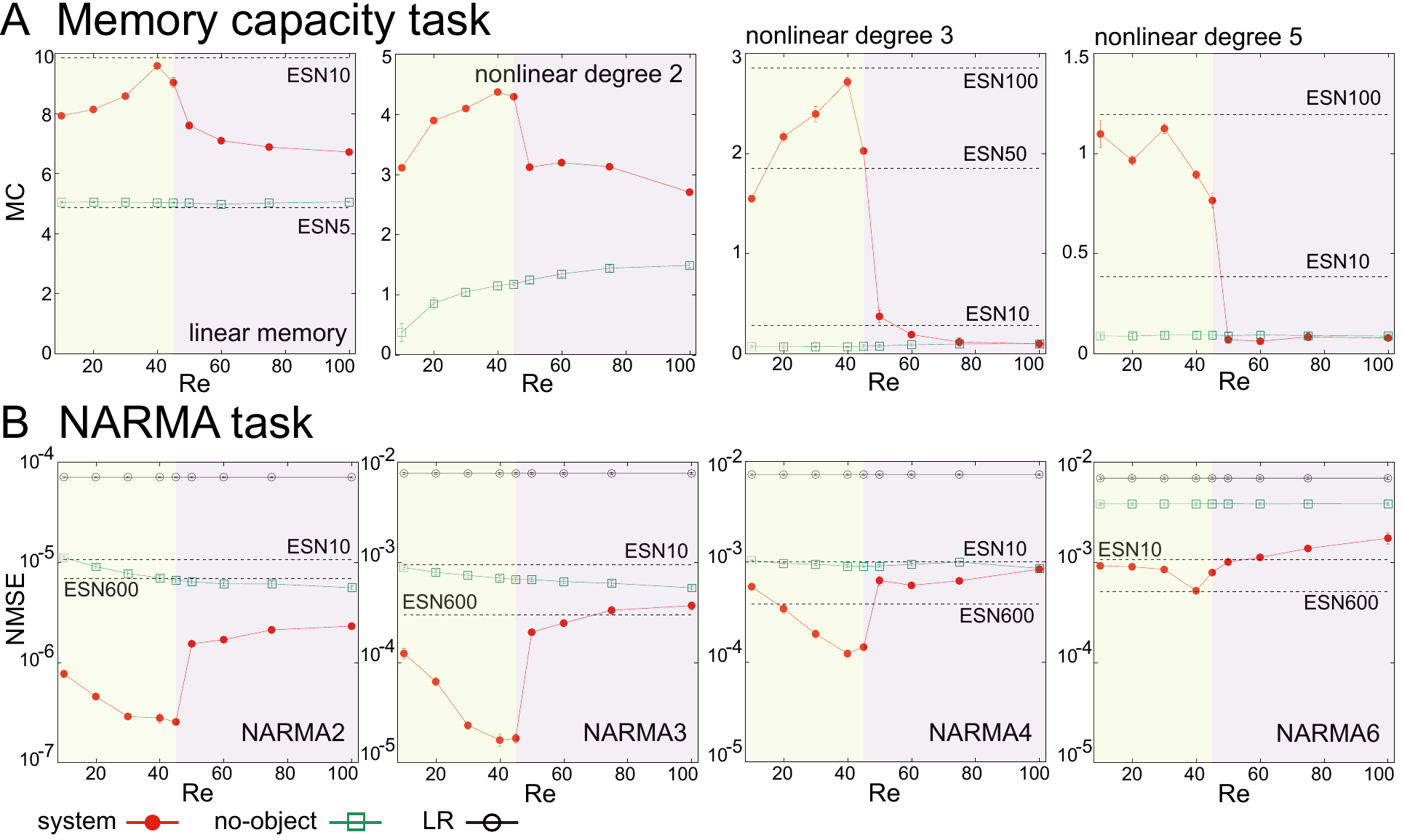}
 \end{center}
 \caption{{\bf Results for evaluations of the memory capacity and NARMA tasks.} ({\bf A}) Results for linear and nonlinear memory capacity are plotted at each Reynolds number. 
({\bf B}) Results for the NARMA tasks in terms of {\it NMSE} are plotted at the Reynolds number of $Re=10,20,30,40,45,50,60,75,$ and $100$. In both A. and B., the plots and its error bar show the mean and its standard deviation, respectively. ESN5, ESN10, ESN50, ESN100, and ESN600 mean $N_{\rm ESN}=5,10,50,100,$ and $600$, respectively.
We clarify the involvement of the vortex in the computational capability by comparing LR, no-object and ESN with those of the system. We determine that the peak of the computational performance is approximately $Re=40$, immediately before the bifurcation point. The background color shading indicates whether the flow is stable (yellow) or unstable (purple).  }
 \label{fig3}
\end{figure*}
Here, we evaluate the memory capacity of the system by investigating whether the system can reproduce previous inputs and nonlinearly process them using its current states. In this study, we apply the Legendre polynomials for each time step expressed as $y_{k}=P_{n}(z_{k-d})$, and $d$ represents the delay gives a value,  $d=0,1,\ldots,50$:
\begin{align}
P_n(l)=\frac{1}{2^n n!}\frac{d^n}{dl^n}[(l^2-1)^n]. 
\end{align} 
 The finite products of the Legendre polynomials for each time step were used in {\it\cite{Dambre-Verstra}}; however, for simplicity,  the nonlinear degree of the Legendre polynomials for each delay is changed and treated as a target, e.g., if $n=3$, $P_3=\frac{1}{2}(5z_{k-d}^3-3z_{k-d})$. In this task, we use the memory function and memory capacity as follows:
 \begin{align}
{\it MF} (d) &\coloneqq\frac{cov^2(y_{k},\hat y_{k})}{\sigma^2(y_{k})\sigma^2(\hat y_{k})}, \\
{\it MC} & \coloneqq \sum_{d=1}^{\infty} {\it MF}(d),
\end{align}
where $\sigma^2(y_k)$ represents the variance of $y_k$.
The objective of  this task is to quantify whether the system can regenerate previous inputs and whether the system can emulate the nonlinear functions of the previous inputs. The quantified values are referred to as the linear memory capacity and nonlinear memory capacity, respectively. 
The memory capacity shows a high value if the system outputs successfully emulate the target outputs for each target delay. Fig.~\ref{fig3}A shows the results of {\it MC} in the evaluation phase. The most striking feature is that {\it MC} increases as the system approaches the critical Reynolds number and suddenly decreases if the Reynolds number exceeds the critical value. In other words, {\it MC} increases as the state of the dynamics approaches from the twin vortex to the Karman vortex and significantly decreases after the transition. At $Re=40$, immediately before the bifurcation point, the system has the highest memory, with ${\it MC}_{Re=40}\approx 9.579$ as the actual value of linear memory.
Moreover, the extensive twin vortices have higher input-sensitivity (cf. Fig.~\ref{fig2}F) and we infer that implementation of the higher computational capability occurs via higher input-sensitivity.
We confirm that the highest {\it MC} in the system shows a similar value to the results of ESN10 for linear memory, while the system performance is comparable to ESN100 in case of nonlinear memory.  Furthermore, our system can represent both odd functions and even functions, whereas some reservoirs, such as ESN, can only demonstrate either of the functions because of the constraint of the setting of the activation functions{\it\cite{Dambre-Verstra}}.

\subsection*{NARMA task}
\begin{figure}[!t]
 \begin{center}
  \includegraphics[width=89mm]{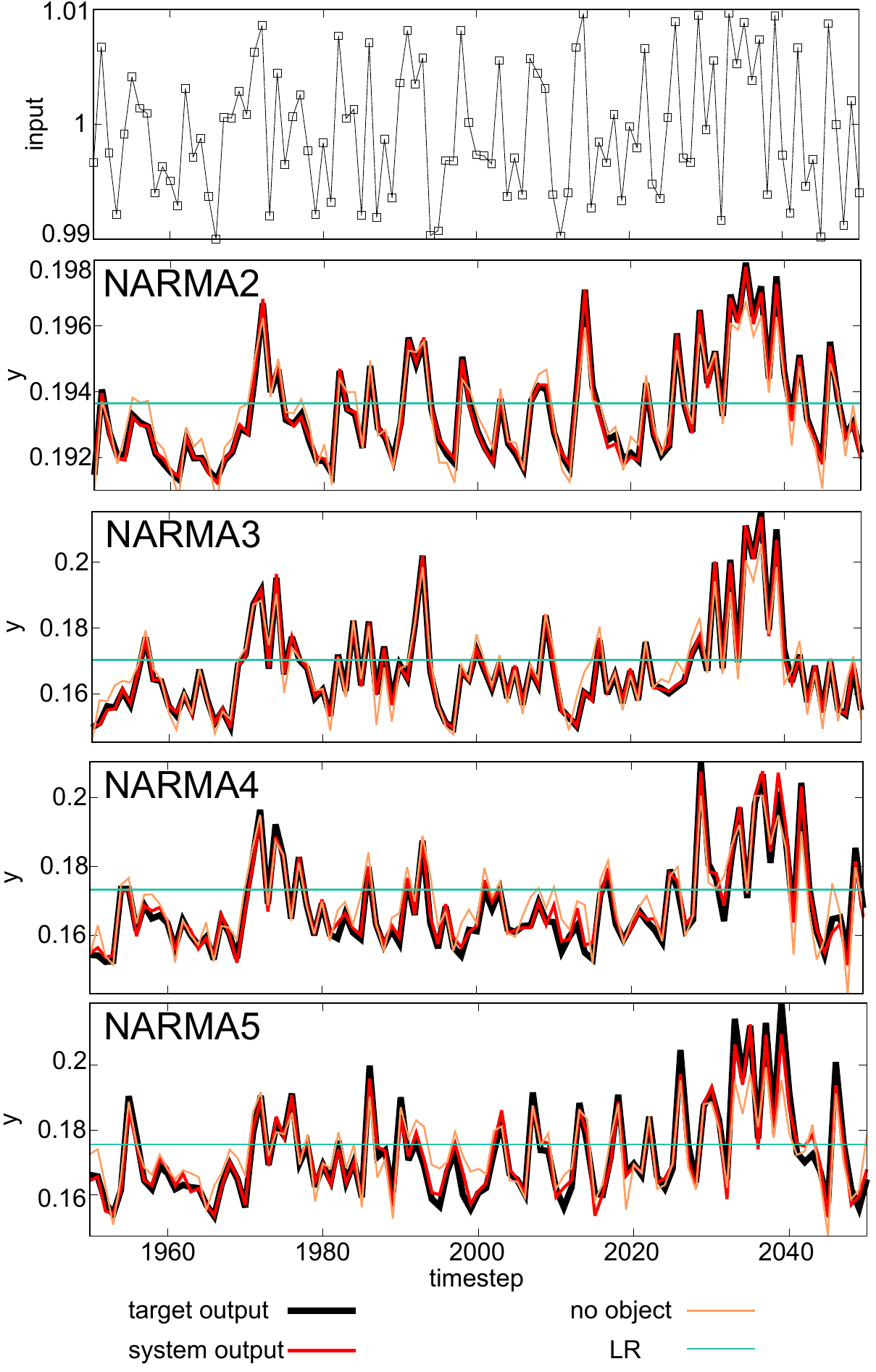}
 \end{center}
 \caption{{\bf A typical performance of the NARMA tasks in the evaluation phase at $Re=40$.} In each plot, the performance of the LR system, system, and no-object are compared with the target. We confirm that the system can trace accurately through the involvement of a vortex.}
 \label{fig4}
\end{figure}
The NARMA model was developed by Atiya and Parlos{\it\cite{Atiya-Parlos}}, and the objective of this task is to model the state of a system depending on an input and its history in a dynamic system with strong nonlinearity; i.e., an emulation of the nonlinear dynamic system. The essential point of the NARMA task is to include the long-term dependencies of the system with $n$-th time lag. We initially introduce the NARMA model of a second-order nonlinear dynamical system as follows:
\begin{align}
  y_{k+1} = 0.4y_{k}+0.4y_{k}y_{k-1}+0.6z_{k}^3+0.1.
\end{align}
In this study, we call this system NARMA2 and the NARMA system such that the $n$-th order nonlinear dynamical system can be written as follows:
\begin{align}
  y_{k+1} = 0.3y_{k} + 0.05y_{k} \biggl( \sum_{j=0}^{n-1} y_{k-j} \biggr) \notag\\
  + 1.5 z_{k-n+1}z_{k}+0.1.
\end{align}
Similarly, these tasks are called NARMA3, NARMA4, NARMA5, and so forth. 
\par
Fig.~\ref{fig3}B shows the error evaluation in the evaluation phase and indicates the averaged normalized mean squared error (NMSE) in each NARMA task separately based on each Reynolds number; NMSE is calculated as ${\it NMSE} \coloneqq \sum_{k}(y_k-\hat y_k)^2/\sum_{k}y_k^2$.
Here, the system, no-object, ESN, which is the standard recurrent neural network, and LR are compared in terms of {\it NMSE} and plotted for each Reynolds number to demonstrate the characteristics of the computational performance of a system (details for the setting of ESN are provided in SM). 
We compare the system with no-object and observe that vortices work effectively depending on the task difficulty and the Reynolds number, and compare the system based on ESN or LR and observe that the fluid dynamics work effectively.
In the results, we observe that the Reynolds number at the optimal performance coincides with the critical Reynolds number of the Karman vortex system (Fig.~\ref{fig3}B). In NARMA2, NARMA3, and NARMA4,  system performance at $Re=40$ is much better than the result of ESN with 600 nodes. For example, the system performances at $Re = 10$ and $100$ are inferior to the performance at $Re = 40$ (Fig.~\ref{fig3} and the supplementary videos).
The plots in Fig.~\ref{fig4} are the representative examples of the actual performances of NARMA2, NARMA3, NARMA4, and NARMA5 at $Re=40$ and demonstrate that the system can clearly trace the target model when compared with the LR system (the supplementary videos). We confirm that tracking accuracy is improved by the involvement of a vortex in comparison with that observed in the no-object case.
We can infer that the nonlinear processing capability is low at $Re = 10$ because the input-sensitivity of vortices is low even though the synchronization occurred and that the computational capability is low at $Re = 100$ because the synchronization is broken.

\subsection*{Time series predictions}
\begin{figure*}[!t]
 \begin{center}
  \includegraphics[width=183mm]{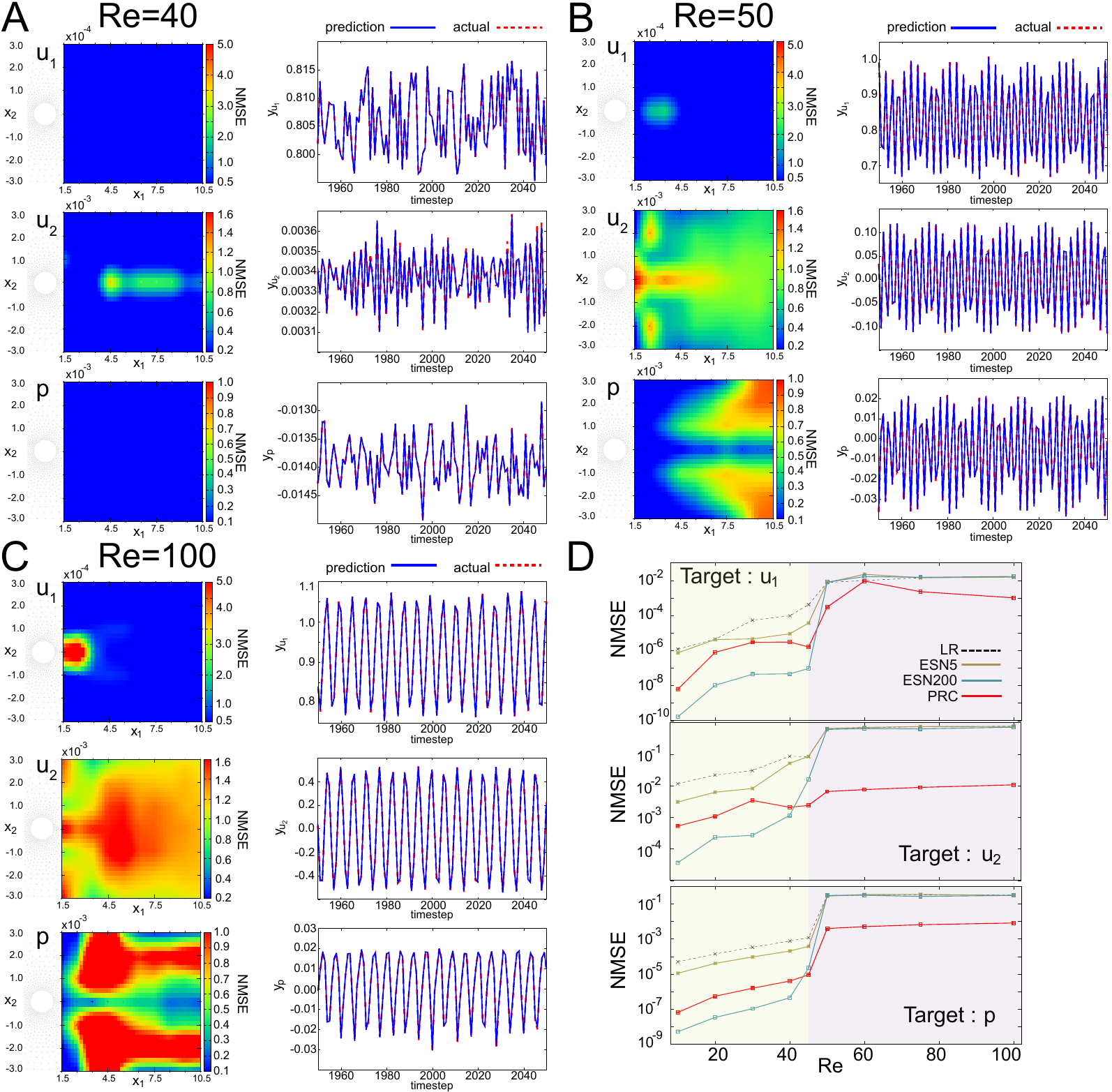}
 \end{center}
 \caption{{\bf Results for the time series prediction task.} The color map shows {\it NMSE} at each the target, namely, $u_1$, $u_2$, and $p$ in the area $[1.5, 10.5] \times [-3, 3]$, and the diagram next to each color map plots the actual performance at typical coordinates by target. We describe $y_{u_1}$ if $y_k=u_1(P_\ell,k\tau)$, $y_{u_2}$ if $y_k=u_2(P_\ell,k\tau)$, and $y_p$ if $y_k=p(P_\ell,k\tau)$. The actual performances of $u_1$, $u_2$, and $p$ are plotted by $(x_1,x_2)=(2.5, 0.0), (4.5, 0.0),$ and $(10.5, -2.0)$, respectively. The Reynolds numbers of ({\bf A}), ({\bf B}), and ({\bf C}) are $Re=40,50,$ and $100$, respectively. ({\bf D}) shows the mean plotted for {\it NMSE} vs. $Re$ for each variable by target. All the nodes in the reservoir region are predicted and the $\ell^2$ norm of the error is calculated. We also conduct a comparison with the LR system and ESN.
}
\label{fig5}
\end{figure*}
RC enables highly accurate predictions of time series data and can estimate the Lyapunov exponents of high dimensional spatio-temporal chaotic systems{\it\cite{Pathak-Lu,Lu-Pathak}}. The behavior of fluid variables is predicted using RC{\it\cite{Nakai-Saiki}}. This section aims to implement the prediction of fluid variables (velocity and pressure) using the system's own dynamics based on the framework of PRC. The dynamics other than the variable to be predicted are used as the reservoir to predict the unknown variable. Thus, if the target output is $u_1$, such as, $y_k=u_1(P_\ell, k\tau)$, the reservoir $\{x_k^i \in \mathbb{R};\ i=1, \ldots, N_{rr},\ k=1, \ldots, T_r \}$ is defined by
\begin{align*}
x_k^{2(\ell-1)+1} & \coloneqq u_2 (P_\ell, (k-1)\tau+\Delta t), \\
x_k^{2(\ell-1)+2} & \coloneqq p (P_\ell, (k-1)\tau+\Delta t), 
\end{align*}
for $\ell = 1, \ldots, S$, $N_{rr}\coloneqq 2S=154$.
\par
Fig.~\ref{fig5} depicts a typical example of an error, with the target output and actual performance for each target. 
The color map in Fig.~\ref{fig5} shows that the errors increase as the Reynolds number increases; however, the actual performance successfully overlaps the target (Fig.~\ref{fig5}A,B,C).
As we observe the results in Fig.~\ref{fig5}D for each target, although $u_2$ is moderate, the error increases according to the Reynolds number; when the Reynolds number exceeds a critical Reynolds number; moreover, the errors of $u_1$ and $p$ remarkably increase. This observation indicates that the prediction under a condition in which the Karman vortex is generated is relatively more difficult than that under which the twin vortex is generated. 
In Fig.~\ref{fig5}D, the time series prediction of twin vortex is easy for any case. Furthermore, the time series prediction of the Karman vortex is difficult in LR and ESN but is comparatively possible using the PRC system. 
The ESP of the system is collapsed in the region $Re \ge 50$. If the reservoir does not satisfy ESP, its dynamics cannot be uniquely determined by the input signal. Therefore, when we reconnect the test data with the optimal weights in the evaluation phase, the error increases.
We can infer that the prediction performance is degraded in this Reynolds number region because of flow instability as the performance of the externally connected ESN model is significantly degraded. 
Moreover, in the aforementioned Memory capacity and NARMA tasks, the computational performance may decrease in this Reynolds number region, and we can consider that the prediction performance using PRC decreases.
In this region $Re \ge 50$, the ESN does not work, and there is less prediction capability using other variables; however, the system has higher prediction capability than that observed using the ESN that constructs an input-only function because the target variable is both a function of inputs and a function of other variables, which contain some of the target variable components, including its history.

\section*{DISCUSSION}
In this study, we first investigate the bifurcation of our system and demonstrate that the Hopf bifurcation occurs in the Karman vortex system with inputs, followed by the synchronization phenomenon in the Karman vortex system, in which the details for the input-sensitivity are investigated. Then, we  implement the PRC framework numerically by using the dynamics of the Karman vortex system. Furthermore, the prediction of a fluid variable is performed by the approach using the dynamics of other variables as the reservoir.
\par
We investigate ESP in terms of direct calculation using reservoir dynamics and stability analysis and determine that the synchronization phenomenon gradually collapses as the phase shift from a twin vortex to a Karman vortex. However, we note that this correspondence between the stability analysis and the Reynolds number holds true only when the input intensity is low and does not necessarily hold true when we use the high input intensity (cf. SM).
The input-sensitivity of the vortices is investigated and the details for vortices' internal structure satisfying ESP are clarified. Furthermore, we suggest that the input-sensitivity of dynamics is useful for estimating of the computational capability.
Moreover, we demonstrate that the physical dynamics of the Karman vortex system can be used as a resource for temporal machine learning tasks and that the computational performance is maximized in the phase just before the synchronization failure.
Note that the critical Reynolds number from the twin vortex to the Karman vortex is known to exist in the fluid field just before this synchronization collapses, and we clarify that the system possesses the essential parts such as a computational performance in a range of $Re\in[40,50]$. 
Furthermore, we clarify that ESP is essential for estimating the computational performance because ESP collapses despite the non-chaotic behavior in our setup.
Moreover, we demonstrate that in the Reynolds number band where ESP is collapsed, that is, the onset of the Karman vortex generation, if we try to predict a behavior of a missing variable within a system. Thus, using the rest of the variables within the same system is more effective to make predictions than constructing external prediction models from scratch.
\par
Finally, although the computational cost of PDEs was much larger than that of ordinary differential equations, it was inevitable for our analyses and to reveal detailed structures of the vortices. 
Our approach suggests the possibility of determining an optimal set of parameters that give high computational capability for the target PRC using a numerical simulation of PDE if the target physical system is modeled well by PDE.

\section*{MATERIALS AND METHODS}
\subsection*{The setup of our reservoir}
We outline a typical RC that consists of input, reservoir, and output layers{\it\cite{Jaeger-Haas1,maass1,luko}}. The reservoir is typically expressed as a high-dimensional dynamic system driven by a low-dimensional input stream, which acts as a type of temporal and finite kernel facilitating the separation of input states. As a mathematical foundation for RC, any filter (time-invariant operator that maps input sequence to output sequence) having fading memory{\it\cite{boyd}} can be approximated with any desired degree of precision by combining a filterbank with a pointwise separation property and a readout function with a universal approximation property{\it\cite{maass2,maass3}}.
Here, if we outsource the nonlinearity of the readout function to a filterbank, then the system turns out to be a typical RC with a linear and static readout. Our aim in this study is to reveal how the Karman vortex system acts as a reservoir in this respect and to analyze its property systematically. 

We briefly review the framework of RC.
Assume that a reservoir map $\bar{F}: \mathbb{R}^M \times \mathbb{R}^m \to \mathbb{R}^m , m,M \in \mathbb{N}$, and a readout map $\bar{h}: \mathbb{R}^M \to \mathbb{R}$ , an infinite discrete input $\bar{{\bf z}}=(\dots,\bar{{\bf z}}_{-1},\bar{{\bf z}}_{0},\bar{{\bf z}}_{1},\dots) \in (\mathbb{R}^m)^{\mathbb{Z}}$and an output signal $\bar{{\bf y}} \in \mathbb{R}^{\mathbb{Z}}$, then a reservoir state vector $\bar{{\bf x}}_k \in \mathbb{R}^M$ and an output signal are determined by $\bar{{\bf x}}_k=\bar{F}(\bar{{\bf x}}_{k-1},\bar{{\bf z}}_k)$ and $\bar{y}_k=\bar{h}(\bar{{\bf x}}_k)$, respectively. To exploit a reservoir map as a filter, it is preferable to have its state be a function of the previous input sequence, $\bar{x}_{k-1} = E (\bar{z}_{k-1}, \bar{z}_{k-2}, \dots)$, which is related to the concept of ESP explained in detail later. These expressions are a generalization of definitions of reservoir computer, and the readout map $\bar{h}$ is a linear map. 
First, we prepare the reservoir corresponding to the function $\bar{F}$ in our system.
\par
We consider flows past a circular cylinder governed by the two-dimensional Navier--Stokes equations with inputs, and numerically solve it by using the stabilized Lagrange--Galerkin (LG) method{\it\cite{Suli,Notsu-Tabata}}. The details for this problem, this method and the setting of inputs are provided in the supplementary materials (SM).
\par
We also provide the definition of our reservoir using the numerical solution $({\bf u}_h^n, p_h^n)$, where ${\bf u}_h^n$ and $p_h^n$ represent the velocity and the pressure, respectively. Note that $({\bf u}_h^n,p_h^n)$ approximates $({\bf u}(\cdot,n\Delta t),p(\cdot,n\Delta t))$, and we use two parameters, $\Delta t$ and $\tau$, for time in the flow simulation. $\Delta t$ is a time step size in the simulation, as mentioned in the previous paragraph, and $\tau$ is a transient time for the reservoir that determines the timescale of the reservoir. The setup of our reservoir is described here.
Let $\Omega \subset \mathbb{R}^2$ be a bounded domain, and
\begin{align*}
\Omega_{R} \coloneqq \{{\bf x} \in \overline{\Omega};\ x_1 = i + 0.5,\ i=0, 1, \ldots, 10,\ \\
x_2 = -3, -2, \ldots, 2, 3\}
\end{align*}
be a set of points, which is, for simplicity, rewritten as $\Omega_R = \{P_i\}_{i=1}^{S}$ for $S\coloneqq \#\Omega_R = 77$ (cf. Fig.~\ref{fig1}).
Let $\tau > 0$ be a transient time in the flow simulation, $T_r \coloneqq \lfloor T_f/\tau \rfloor$ be the total time in our reservoir, and $N_r \coloneqq 3S = 231$ be the total number of computational nodes in our reservoir.
We introduce a notation $n_k \coloneqq k \tau/\Delta t~(k\in\mathbb{N}\cup\{0\})$ with the relation~$t^{n_k} = k\tau$.
Let $\{({\bf u}^r(P_i,k), p^r(P_i,k))\in \mathbb{R}^2\times\mathbb{R};\ i=1, \ldots, S,\ k=0, \ldots, T_r-1\}$ be a part of the numerical solution defined by
\[
({\bf u}^r(P_i,k), p^r(P_i,k)) \coloneqq ({\bf u}_h^{n_{k+1}}(P_i), p_h^{n_{k+1}}(P_i)),
\]
which corresponds to the $k$-th input,
\[
z_k \coloneqq z_{1}^{n_k} = z_{1}(k\tau),
\]
and is imposed as the Dirichlet boundary condition in the system. Note that, the three values of $({\bf u}_h^n, p_h^n) (P_i)$ are uniquely determined because $u_{h1}^n, u_{h2}^n$, and $p_h^n$ are continuous functions defined in $\overline{\Omega}$ (cf.~SM). Using $({\bf u}^r(P_i,k), p^r(P_i,k))$, we define our reservoir $\{x_k^i \in \mathbb{R};\ i=1, \ldots, N_r,\ k=0, \ldots, T_r-1 \}$ by
\begin{align*}
x_k^{3(\ell-1)+1} & \coloneqq u_1^r (P_\ell, k), \\
x_k^{3(\ell-1)+2} & \coloneqq u_2^r (P_\ell, k), \\
x_k^{3(\ell-1)+3} & \coloneqq p^r (P_\ell, k),
\end{align*}
for $\ell = 1, \ldots, S$.
\par
Then, the output $\hat y_k$ is computed by
\begin{align*}
\hat y_k &= \sum_{i=0}^{N_r} w_{out}^i x_k^i,
\end{align*}
where $x_k^0=1$ is a bias and $w_{out}^i$ is the readout weight of the $i$-th computational node $x_k^i$.
As usual in the framework of RC, the target function, $y_k = f (z_k, z_{k-1}, \ldots)$ for  a given function $f$ is learned by adjusting the linear readout weights.
The ridge regression, known as $L^2$ regularization (cf.{\it\cite{Hastie-Tib}}) is employed to obtain the optimal weights.
The performance of the system output is evaluated by comparing $\hat{y}_k$ with the target output~$y_k$.

\subsection*{Descriptions of an indicator of synchronization and the amplification factor}
The common-signal-induced synchronization is such that the states of two different initial condition reservoirs driven by the same input sequence approach the same value over time. Intuitively, it indicates that the reservoir asymptotically washes away the information related to the initial conditions, i.e., this condition indicates that if a certain input sequence is injected at any time, it will return a same certain response, which expresses the minimum characteristics necessary for the reservoir to implement a reproducible input-output relation and calculation. 
\par
A similar concept to common-signal-induced synchronization is ESP in which the current network state $\bar{r}(t)$ is expressed as a function of only the previous input series $\bar{z}(t)$, independent of the initial value $\bar{r}(0)$; i.e., there exists $E$  such that $\bar{r}(t)=E(...,\bar{z}(t-1),\bar{z}(t))${\it\cite{Jaeger1}}. In this study, we directly investigate the degree of synchronization using two reservoir states corresponding to two identical input sequences with different initial values and using the amplification factor of perturbation{\it\cite{Takemoto-Mizushima}}, where, to get different initial values, different initial input sequences are used. Let ${\bf x}_k$ and $\tilde{{\bf x}}_k (\in \mathbb{R}^{N_r})$ be two reservoir states corresponding to two identical input sequences with different initial values; furthermore, we calculate $E_m$:
\begin{align}
E_m \coloneqq \frac{\|{\bf x}_{k+1}-{\bf x}_k\|_{\ell^2(\Omega_R)}}{\|\tilde{{\bf x}}_{k+1}-\tilde{{\bf x}}_k \|_{\ell^2(\Omega_R)}}
\label{def:Em}
\end{align}
for the norm $\|{\bf x}_k\|_{\ell^2(\Omega_R)} \coloneqq \{ \sum_{i=1}^{N_r} (x_k^i)^2 \}^{1/2}$. $E_m$ asymptotically converges to $1$ if ${\bf x}_k$ and $\tilde{{\bf x}}_k$ systems are synchronized with each other.
\par
Then, the stability analysis of the fluid solution using perturbation investigates whether the perturbation given as an initial value increases or decreases with time. The amplification factor of the perturbation is obtained by numerically solving the perturbation equation described in the SM and we obtain the following $\lambda$ (let $\tilde{{\bf u}}_h = \{ \tilde{{\bf u}}_h\}_{n=0}^{N_T}$ be the perturbation for the velocity obtained by solving the perturbation equation in SM)
\begin{align}
\lambda \coloneqq \frac{1}{n}\log \frac{\| \tilde{{\bf u}}_h^n\|_{L^2(\Omega)}}{\|\tilde{{\bf u}}_h^0\|_{L^2(\Omega)}},
\label{def:ESP}
\end{align}
which is the amplification factor of the perturbation $\tilde{{\bf u}}_h$. If $\lambda>0$, the perturbation increases exponentially, and the solution is unstable in $\Omega$ in this sense. Because this stability analysis is usually performed for steady solutions, we analyze our setup and the steady solution for comparison, i.e., we conduct analysis at $z_{1}\equiv 1$ (called no-input).  Notably, the amplification factor of the perturbation, obtained by stability analysis of the fluid solution and the Lyapunov exponent (an indicator to quantitatively evaluate initial value sensitivity containing a feature of chaos) differ in derivation. 
\subsection*{Computational fluid dynamics}
Let $\Omega \subset \mathbb{R}^2$ be a bounded domain, $\partial\Omega$ the boundary of $\Omega$, and $T_f$ a positive constant.
We suppose that $\partial\Omega$ comprises four parts, $\Gamma_i (\subset \partial\Omega)$, $i=1, \ldots, 4$.
Our problem is to find $({\bf u}, p): \Omega \times (0,T_f)\to \mathbb{R}^2 \times \mathbb{R}$ such that
\footnotesize
\begin{subequations}\label{prob:NS}
\begin{align}
\frac{\partial {\bf u}}{\partial t} + ({\bf u}\cdot\nabla){\bf u}  - \nabla \cdot \sigma({\bf u}, p) & = {\bf 0} && \!\!\!\!
\mbox{in} \ \Omega \times (0,T_f),
\label{NS_eq1} \\
\nabla \cdot {\bf u} &= 0 && \!\!\!\!
\mbox{in} \ \Omega \times (0,T_f), \\
{\bf u} & = {\bf z} && \!\!\!\!
\mbox{on} \ \Gamma_{1} \times (0,T_f), \\
[\sigma({\bf u},p){\bf n}] \cdot {\bf n}^\perp = 0, \ {\bf u}\cdot {\bf n} & = 0 
&& \!\!\!\!
\mbox{on} \ \Gamma_{2} \times (0,T_f), \\
\sigma({\bf u},p){\bf n} & = {\bf 0} && \!\!\!\!
\mbox{on} \ \Gamma_{3} \times (0,T_f), \\
{\bf u} & = {\bf 0} && \!\!\!\!
\mbox{on} \ \Gamma_{4} \times (0,T_f), \\
{\bf u} & = {\bf u}^0 && \!\!\!\!
\mbox{in} \ \Omega,\ \mbox{at} \ t=0,
\end{align}
\end{subequations}
\normalsize
where ${\bf u}=(u_1, u_2)^T$ is the velocity, $p$ is the pressure,
$\sigma({\bf u}, p) := (2/Re) D({\bf u})-pI$ is the stress tensor,
$D({\bf u}) := (1/2)[ \nabla {\bf u} + (\nabla {\bf u})^T ] \in \mathbb{R}^{2 \times 2}$ is the strain-rate tensor,
$I \in \mathbb{R}^{2 \times 2}$ is the identity tensor,
${\bf n}: \partial\Omega \to \mathbb{R}^2$ is the outward unit normal vector,
${\bf n}^\perp: \partial\Omega \to \mathbb{R}^2$ is the unit tangential vector,
${\bf z}: \Gamma_1\times (0, T_f) \to \mathbb{R}^2$ is a given boundary velocity,
and ${\bf u}^0: \Omega \to \mathbb{R}^2$ is a given initial velocity.
\par
In our system we set, for $L\coloneqq 7.5$,
\begin{align*}
\Omega & \coloneqq \{ {\bf x} \in \mathbb{R}^2;\ x_{1} \in (-L, 3L), x_{2} \in (-L, L), |{\bf x}| > 0.5 \}, \\
\Gamma_1 & \coloneqq \{ {\bf x} \in \partial\Omega;\ x_1 = -L, \ x_2 \in [-L, L]  \}, \\
\Gamma_2 & \coloneqq \{ {\bf x} \in \partial\Omega;\ x_1 \in (-L, 3L), \ x_2 \in \{-L, L\}  \}, \\
\Gamma_3 & \coloneqq \{ {\bf x} \in \partial\Omega;\ x_1 = 3L, \ x_2 \in [-L, L]  \}, \\
\Gamma_4 & \coloneqq \{ {\bf x} \in \partial\Omega;\ |{\bf x}| =0.5 \},
\end{align*}
i.e., $\Omega$ is the computational domain, $\Gamma_1$ is an inflow boundary on the left side, $\Gamma_2$ is a slip boundary on the top and bottom side, $\Gamma_3$ is a stress-free boundary on the right, and $\Gamma_4$ is the no-slip boundary on the circle.
We also set
${\bf z} = (z_{1}, 0)^T$ for an {\em input} function $z_{1} = z_{1}(t) \in \mathbb{R}$, where the input $z_1$ takes a random value at each fixed time interval, cf. the subsection of platform setting for details.
The range of $z_{1}$ is set as $[0.99, 1.01]$, and experiments with the other ranges are studied in the SM. 
\par
We solve problem~\eqref{prob:NS} numerically using the stabilized LG method{\it\cite{Suli,Notsu-Tabata}}, a stabilized finite element method combined with the idea of the method of characteristics. Moreover, the material derivative is discretized along the trajectory of a fluid particle as
\footnotesize
\begin{align*}
\Bigl[ \frac{\partial {\bf u}}{\partial t} & + ({\bf u}\cdot \nabla) {\bf u}\Bigr] ({\bf x}, t^n) \\
& \approx \frac{{\bf u}^n({\bf x})-{\bf u}^{n-1}({\bf x}-{\bf u}^{n-1}({\bf x})\Delta t)}{\Delta t},
\end{align*}
\normalsize
where $\Delta t$ is a time increment for the flow computation, and $N_T \coloneqq \lfloor T_f/\Delta t \rfloor$, $t^n \coloneqq n \Delta t$, and ${\bf u}^n \coloneqq {\bf u}(\cdot, t^n)$, $p^n \coloneqq p(\cdot, t^n)$ for $n \in \{ 0, \ldots, N_T \}$.
Let $h>0$ be a representative mesh size.
The numerical solution to be obtained by the LG method is a series of piecewise linear continuous functions defined in~$\Omega$, $\{({\bf u}_h^n, p_h^n):\ \Omega \to \mathbb{R}^2\times\mathbb{R};\ n=1,\ldots, N_T \}$.
Note that, $({\bf u}_h^n, p_h^n)$ approximates $({\bf u}^n, p^n)$.
For the fully discretized scheme, please refer to scheme in SM of LG method section.

\renewcommand{\refname}{References}

\bigskip\bigskip
\noindent
{\bf Acknowledgments:}\ 
This work is supported by JST PRESTO Grant Number~JPMJPR16EA, and JSPS KAKENHI Grant Number JP18H01135. This work is partially based on results obtained from a project commissioned by the New Energy and Industrial Technology Development Organization (NEDO).

\end{document}


\title{SUPPLEMENTARY MATERIALS:\\
Computing with vortices:
Bridging fluid dynamics and its information-processing capability
}
\author{Ken Goto}
\affiliation{Division of Mathematical and Physical Sciences,
Kanazawa University, Kakuma, Kanazawa 920-1192, Japan. \\}%

\author{Kohei Nakajima}
\altaffiliation{Corresponding author. \\
 {\quad} k\_nakajima@mech.t.u-tokyo.ac.jp\\
 }
\affiliation{Graduate School of Information Science and Technology, The University of Tokyo, Bunkyo-ku, 113-8656 Tokyo, Japan.\\}

\author{Hirofumi Notsu}
\affiliation{Faculty of Mathematics and Physics, Kanazawa University, Kakuma, Kanazawa  920-1192, Japan. \\}
\affiliation{PRESTO, Japan Science and Technology Agency, Kawaguchi, Saitama 332-0012, Japan.\\}

\maketitle
\clearpage
\section*{Platform setting and other analyses under some conditions}
We numerically solve our Navier--Stokes problem with input~$z_{1}$ to use the system of the Karman vortex as the reservoir. We provide the complete definition of the series $\{z_{1}^n\}_{n=1}^{N_T} \subset \mathbb{R}$, which is provided by using random numbers.
\par
Let $\Delta t$ be a time increment, and $\tau$ a transient time for each input.
Let $k_\ast \coloneqq \lfloor T_f/\tau \rfloor \in\mathbb{N}$, $n_k\coloneqq k \tau/\Delta t~(k=0,\ldots,k_\ast)$ with the relation~$t^{n_k} = k \tau$, and
\begin{align*}
{\rm sgn}(a) \coloneqq
\left\{
\begin{aligned}
& \frac{a}{|a|} && ( a \neq 0 ), \\
& 0 && ( a = 0 ).
\end{aligned}
\right.
\end{align*}
\par
Let a series of input~$\{r_k\}_{k = 0}^{k_\ast} \subset [ \underline{r}, \overline{r}]$ be given, where $\underline{r} \coloneqq 1-\rho, \overline{r} \coloneqq 1+\rho \in \mathbb{R}$ for an {\em intensity}~$\rho \ge 0$, $[\underline{r}, \overline{r}]$ is the range of random numbers, for example, $[\underline{r}, \overline{r}] = [0.99,1.01]~(\rho=0.01), [0.5,1.5]~(\rho=0.5)$, and the expectation value (probability average) of $\{r_k\}_{k = 0}^{k_\ast}$ is $1$.
For $t\in (t^{n_k}, t^{n_{k+1}}] = (k\tau, (k+1)\tau]$, we introduce a notation~$\widetilde{r}_{k+1}(t)$ defined by
\[
\widetilde{r}_{k+1}(t) \coloneqq r_k + {\rm sgn}(r_{k+1}-r_k) \, \alpha \, (t-t^{n_k}),
\]
where $\alpha \ge 0$ is a tilt parameter, and $\widetilde{r}_{k+1}$ shows a value in progress of the transition from~$r_k$ to~$r_{k+1}$.
Now, we give the definition of the function $z_{1}: (0, T_f) \to \mathbb{R}$ used in our Navier--Stokes problem as, for $t\in (t^{n_k}, t^{n_{k+1}}]$ and $k\in\{0,\ldots,k_\ast-1\}$,
\begin{align*}
z_{1}(t)
&\coloneqq 
\left\{
\begin{aligned}
& \widetilde{r}_{k+1}(t) && \bigl( t^{n_k} < t < t^{n_k} + \tfrac{|r_{k+1} - r_k|}{\alpha}\bigr), \\
& r_{k+1} && \bigl( t^{n_k} + \tfrac{|r_{k+1} - r_k|}{\alpha} \le t \le t^{n_{k+1}} \bigr),
\end{aligned}
\right.
\end{align*}
which provides $\{z_{1}^n\}_{n=1}^{N_T}$ explicitly.
\par
The parameter values used in the computation are summarized in Table~1.
We numerically solve our Navier--Stokes problem for a time period $(0, T_f)$ with $T_f = (T_{ii} + T_{i})\tau = 10500$, which implies that the total number of time steps $N_T$ for the fluid computation is $N_T = (T_{ii} + T_{i}) \tau/\Delta t = 52500$, where $T_i$ is a number of inputs to be used for the reservoir, and $T_{ii}$ is a number of initial inputs not to be used for it.
\par
First, we must know the behavior of flow past a circular cylinder with input, because, according to our review of the literature, no researchers had investigated the flow past a circular cylinder with a (time-dependent) input corresponding to our computation; however, we did observe a substantial amount of literature on the flow without input, that is, $z_{1}^n=1$ for all $n\in\mathbb{N}~(n \le N_T)$. In the following, we call the flow with $z_{1}^n=1$ the no-input system.
\par
Based on numerical experiments, the critical Reynolds number~$Re_c$ of the onset of (wake) instability of flow in the no-input system is in a range $Re_c \in [40, 50]$ (cf. e.g.,{\it\cite{Takemot-Mizushima}} and the references therein).
In the case of our system, however, the flow behavior depends on the range of $u_{\ast1}^n$. Notably, in the case $[\underline{r}, \overline{r}] = [0, 2]~(\rho = 1)$, our system generates vortices even for $Re=30$.
\par
We observe the flows in our system with $Re\in [10, 100]$ and four cases of the range of $u_{\ast1}$, $[\underline{r}, \overline{r}] = [0, 2]~(\rho = 1), [0.5, 1.5]~(\rho = 0.5), [0.9, 1.1]~(\rho = 0.1),$ and $\{1\}$~($\rho = 0$, the no-input system), by computing the lift coefficient~$C_L$ defined by
\begin{align*}
C_L^n = C_L(t^n) \coloneqq -2\int_{\Gamma_4}[\sigma({\bf u}_h^n,p_h^n) {\bf n}]_2 \, ds.
\end{align*}
\par
The graphs of $C_L(t)$ with respect to $t$ for $Re \in [Re_c, 100]$ are smooth and periodic curves, which implies that the flows are non-symmetric with respect to the $x_1$-axis.
We employ the discrete Fourier power spectrum for the analysis of $C_L^n$ as follows:
\begin{align*}
C_L^n \approx \sum_{k=0}^N \Bigl[ a(k)\cos\frac{2\pi kn}{N} +{\rm j} b(k)\sin\frac{2\pi kn}{N} \Bigr],
\end{align*}
where $\rm j$ is the imaginary unit, $N=2^{15}~(\le N_T)$ is the number of data used for the analysis, and $a(k)$ and $b(k)$ are the so-called Fourier coefficients.
\par
\begin{figure}[!t]
\begin{center}
\includegraphics[width=89mm]{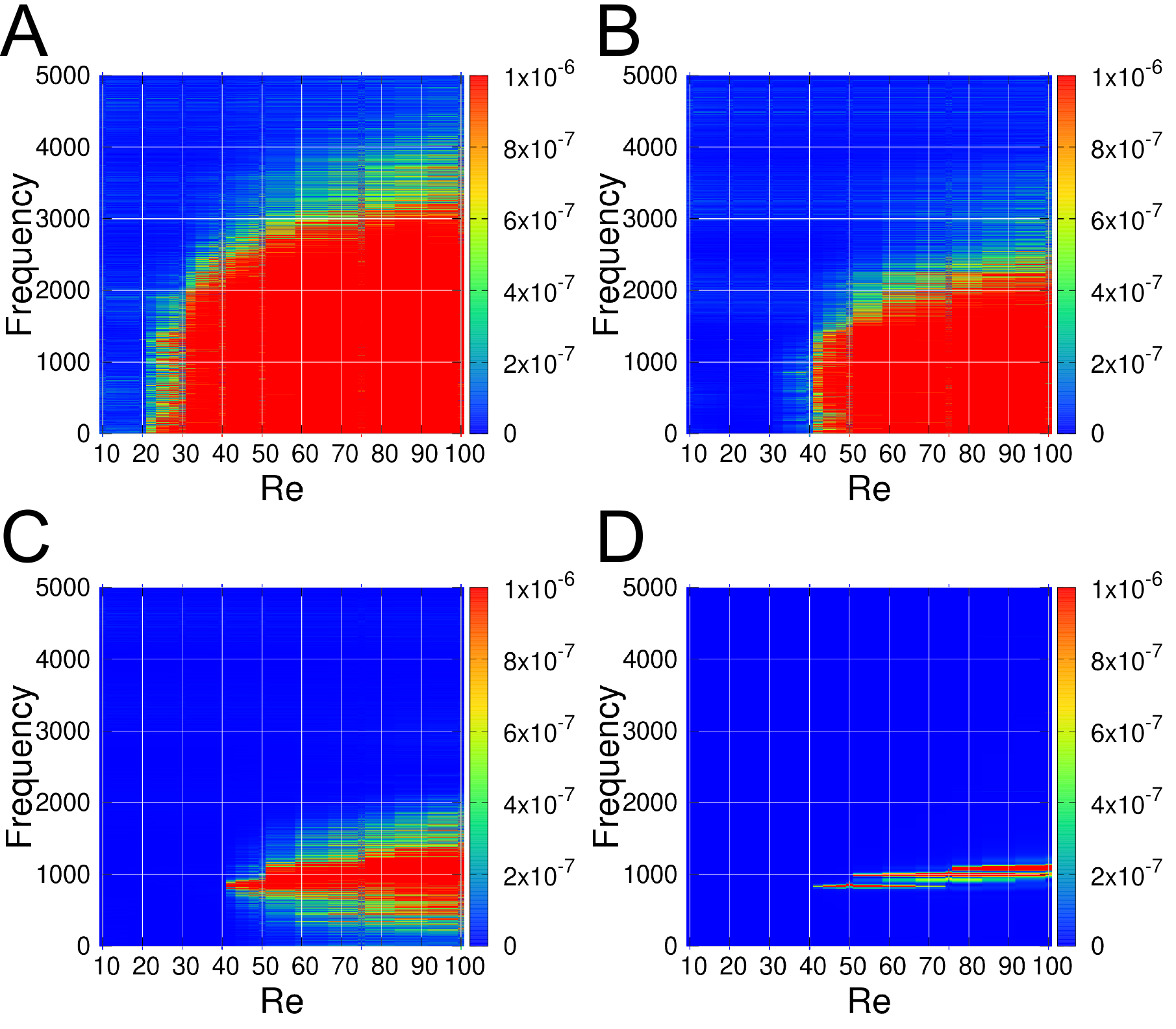}
\end{center}
\caption{{\bf Power spectrum with respect to frequency and Reynolds number by the discrete Fourier transform of lift coefficient~$C_L^n$.}
({\bf A}), ({\bf B}), ({\bf C}), and ({\bf D}) correspond to the results with
$[\underline{r}, \overline{r}] = [0, 2]~(\rho = 1), [0.5, 1.5]~(\rho = 0.5), [0.9, 1.1]~(\rho = 0.1),$ and $\{1\}$~($\rho = 0$, the no-input system), respectively, where the computation is performed for $Re=10, 20, 30, 40, 50, 75, $and $100$. The spectrum values between two Reynolds numbers are interpolated.}
\label{fig6}
\end{figure}

Fig.~\ref{fig6} shows the color map of the discrete Fourier power spectrum, that is, a power spectrum with respect to frequency and Reynolds number by the discrete Fourier transform of~$C_L^n$.
Typical results of the no-input system are obtained in D ($[\underline{r}, \overline{r}] = \{1\}$, $\rho=0$), and a similar trend to D is shown in C ($[\underline{r}, \overline{r}] = [0.9, 1.1]$, $\rho=0.1$).
Furthermore, in the case that the range of $u_{\ast1}$ is $[\underline{r}, \overline{r}] = [0.99, 1.01]$~($\rho=0.01$) used in the main text, we obtain that the spectrum values are similar to values of D, the no-input system.
\par
In A ($[\underline{r}, \overline{r}]=[0, 2], \rho=1$) and B ($[\underline{r}, \overline{r}]=[0.5, 1.5], \rho=0.5$), however, we can observe results different from C and D.
When the range of $u_{\ast1}^n$ is wide, that is, a {\em high intensity} case, the spectrum values are high at low Reynolds numbers, $Re=30, 40$.
In other words, the typical spectrum values of the no-input system can be observed, for example, at $Re=30$ if the intensity is high.
\par
\begin{figure*}[!t]
\begin{center}
\includegraphics[width=183mm]{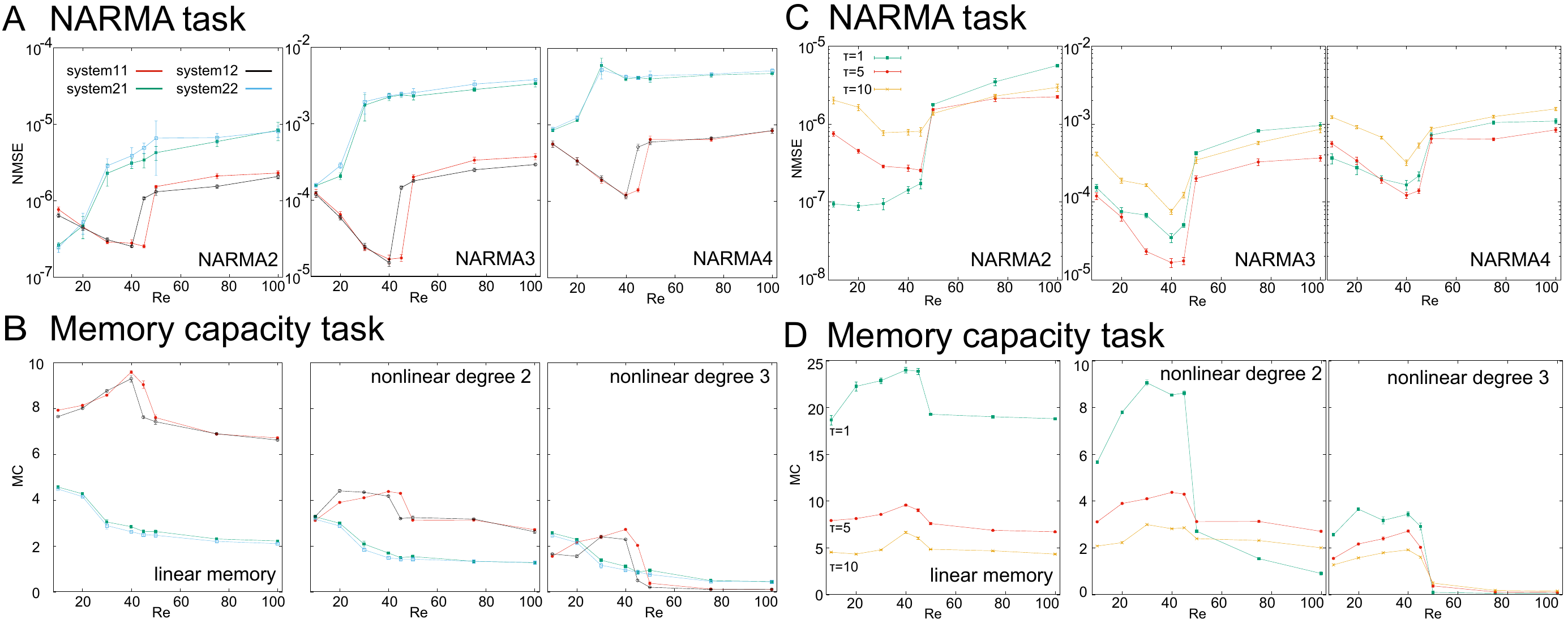}
\end{center}
\caption{{\bf The task results by two mesh, or by transient time parameters.} ({\bf A}) and ({\bf B}) are the mean and deviation of results from the NARMA task and the evaluation of memory capacity, respectively, where
system11 is obtained by using range $[\underline{r}, \overline{r}] = [0.99, 1.01]$ and Mesh~1; 
system12 is obtained by using range $[\underline{r}, \overline{r}] = [0.99, 1.01]$ and Mesh~2; 
system21 is obtained by using range $[\underline{r}, \overline{r}] = [0, 2]$ and Mesh~1; 
and system22 is obtained by using range $[\underline{r}, \overline{r}] = [0, 2]$ and Mesh~2.
The results imply that the mesh dependency is not strong. ({\bf C}) and ({\bf D}) are the mean and deviation of results from the two tasks, where the results are obtained by a transient time parameter $\tau=1,5,10$ on system11.}
\label{fig7}
\end{figure*}
\begin{figure}[!t]
\begin{center}
\includegraphics[width=89mm]{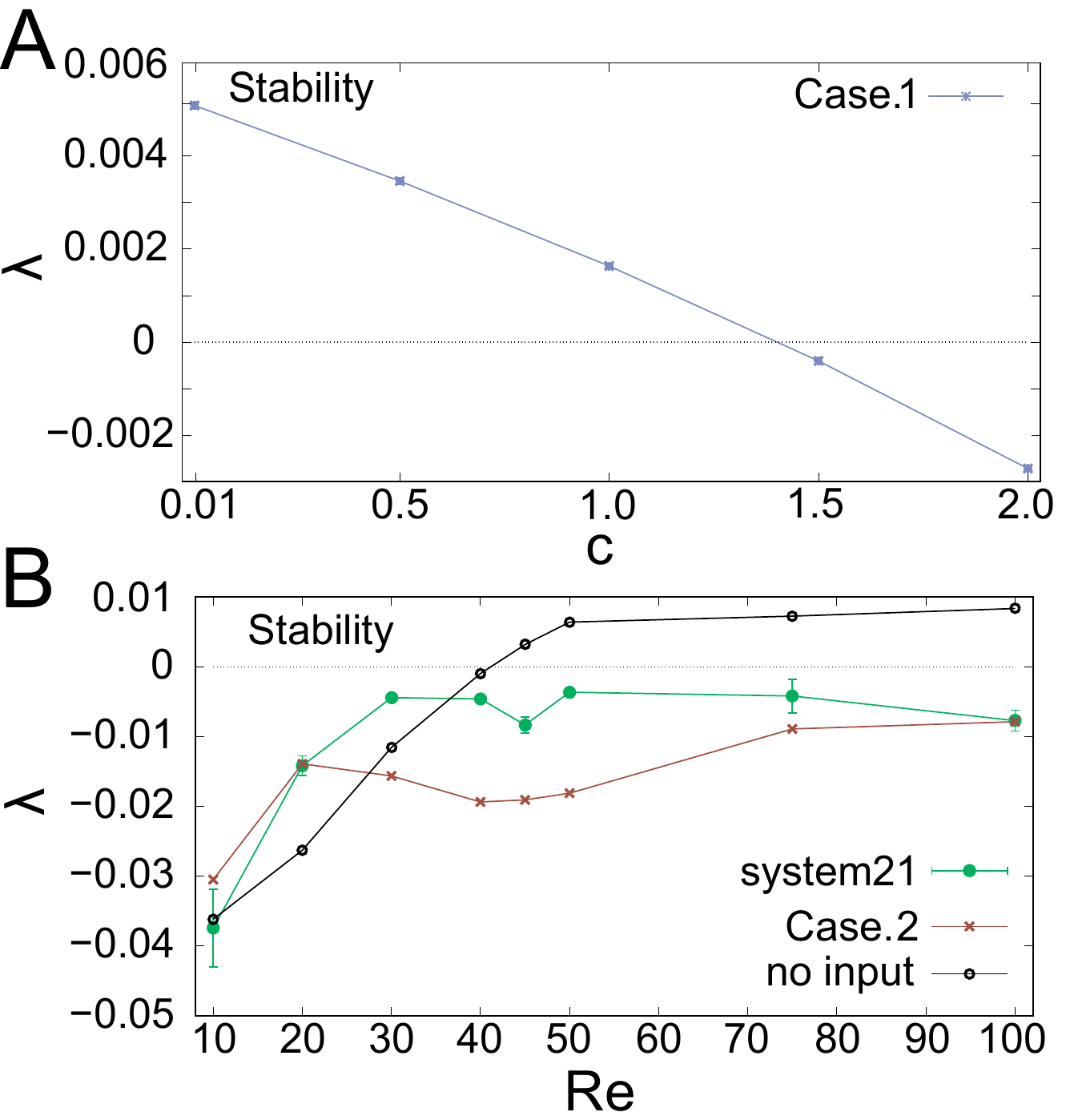}
\end{center}
\caption{{\bf Results of $\lambda$ for some cases.} ({\bf A}) shows the results of $\lambda$ for Case~1 with respect to the constant parameter $c$, where the Reynolds number is fixed at $Re=100$. ({\bf B}) shows graphs of $\lambda$ vs. $Re$ for system21, Case~2, and no-input system.}
\label{fig8}
\end{figure}

Second, we confirm the mesh dependency of the results presented in the main text.
We have employed a mesh with \#nodes, $N_p = 2734$, and \#elements, $N_e=5278$~(Mesh~1) in the main text.
Using a finer mesh with $N_p=6061$ and $N_e=11387$~(Mesh~2), we additionally compute some of the same problems int the NARMA task and Memory capacity, where \#nodes and \#elements of Mesh~2 are more than twice \#nodes and \#elements of Mesh~1.
The results are shown in Fig.~\ref{fig7}A,B, and we describe system11 ($[\underline{r}, \overline{r}]=[0.99, 1.01]$, Mesh~1) and system12 ($[\underline{r}, \overline{r}]=[0.99, 1.01]$, Mesh~2), and we can observe similar results in both A~(NARMA task) and B~(Memory capacity).
The range $[\underline{r}, \overline{r}] = [0,2]$ is also employed in the computation.
Although both performances obtained by using Meshes~1 and~2 for $[\underline{r}, \overline{r}] = [0,2]$ are not high, they are similar, and we describe system21 ($[\underline{r}, \overline{r}]=[0, 2]$, Mesh~1) and system22 ($[\underline{r}, \overline{r}]=[0, 2]$, Mesh~2).
From these results, we conclude that the dependency on the mesh is not strong.
In addition, the transient time parameter $\tau$, which is fixed in the maintext, is changed to $\tau=1,10$ (Fig.~\ref{fig7}C,D).
%
Third, we add the information of results obtained by using a different range~$[\underline{r}, \overline{r}]=[0, 2]~(\rho=1)$.
Hereafter, our system with the range $[\underline{r}, \overline{r}] = [0, 2]$ is called our system with high intensity. We have computed $\lambda$, and, as a result, our system with high intensity has the dynamics synchronization ($\lambda<0$) for all Reynolds numbers less than $100$ (Fig.~\ref{fig8}B), that is, $Re = 10, 20, 30, 40,45, 50, 75,$ and $100$, although the task performance is relatively not good (Fig.~\ref{fig7}A,B).
Notably, our system with high intensity generates Karman vortex at $Re=30$.
\par
Our system with high intensity has dynamics synchronization and generates vortices, whereas the no-input system does not have dynamics synchronization and generates vortices, for example, at $Re=100$.
The results imply that the high intensity stabilizes the system.
To understand this phenomenon, we consider two cases of problems and solve them numerically.
We modify our Navier--Stokes problem by putting an external force~${\bf f}: \Omega\times (0, T) \to \mathbb{R}^2$ into the right hand side of our Navier--Stokes equation, that is, the first equation becomes
\footnotesize
\begin{align}
\frac{\partial{\bf u}}{\partial t} + ({\bf u} \cdot \nabla){\bf u} - \nabla \cdot \sigma({\bf u}, p) &= {\bf f} && \mbox{in}\hspace{5pt}\Omega \times (0,T_f).
\label{NS_eq1_modified}
\end{align}
\normalsize
Our Navier--Stokes problem with equation~\eqref{NS_eq1_modified} instead of our Navier--Stokes equation is simply called problem~\eqref{NS_eq1_modified}.
The first case, Case~1, is to solve problem~\eqref{NS_eq1_modified} with $f_1(t)=c [1 + \sin ( \frac{\pi}{5} t) ]$, $f_2 = 0 $, $z_{\ast1} = 1$, and $Re=100$, where $c\in\mathbb{R}$ is a parameter.
The second case, Case~2, is to solve problem~\eqref{NS_eq1_modified} with ${\bf f} = {\bf 0}$, $z_{\ast1} (t) = 1 + \sin (\frac{\pi}{5} t)$, and $Re=10, 20, 30, 40, 45, 50, 75,$ and $100$.
Case~1 has an external force and no-input, and Case~2 has no external force and an input whose range is $[0, 2]$.
We compute Cases~1 and 2, where five values of $c$, that is, $c=0.01, 0.5, 1.0, 1.5,$ and $2.0$, are employed for Case~1.
The results are shown in Fig.~\ref{fig8}, A shows that the high value of $c$, that is, the large external force, stabilizes the solution at $Re=100$, which implies instability in the no-input system.
As shown in B, the solutions to both system21 and our system with a similar input $z_{\ast 1}(t)=1+\sin(\frac{\pi}{5}t)$ are stable even for the cases $Re \ge 50$, which generate vortices in the no-input system.
From these results shown in Figs.~\ref{fig7} and~\ref{fig8}, we observe that a high input intensity (or a large external force) generates vortices, stabilizes the system, and leads to dynamics synchronization. The system, however, has the low computational performance. Further analysis on the input intensity is planned for further research.
 
\section*{Lagrange--Galerkin method}
In this subsection, the LG method employed in this paper is explained.
\par
Let $\mathscr{T}_h := \{K_{\ell}\}_{\ell=1}^{N_e}$ be a triangulation of the domain~$\Omega$, 
where
$K_\ell$ is a (closed) triangular element, 
$N_e$ is a total number of elements,
$h$ is a representative mesh size, and 
$\Omega_h := {\rm int}(\bigcup^{N_e}_{\ell=1} K_{\ell})$ is an approximate domain of~$\Omega$.
Although $\Omega_h \neq \Omega$ holds in our system setting due to the presence of curved boundary, we do not use the notation~$\Omega_h$ in the following and do not introduce new notations for discrete boundaries in order to avoid unnecessary confusion.
For a function ${\bf g}: \Gamma_{1} \to \mathbb{R}^2$ let $M_h$, $V_h({\bf g})$, $V_h$, and $Q_h$ be function spaces defined by
\begin{align*}
M_h &:= \{q_h \in C(\overline{\Omega});\ q_{h|K}\in \mathscr{P}_1(K), \forall K \in \mathscr{T}_h\}, \\
V_h( {\bf g} ) &:= \{ {\bf v}_h \in M_h^2;\ 
{\bf v}_{h|\Gamma_{1}} = {\bf g},\
({\bf v}_h \cdot {\bf n})_{|\Gamma_{2}} = 0,\\
& \qquad {\bf v}_{h|\Gamma_{4}} = {\bf 0} \},\\
V_h &:= V_h(0),\quad
Q_h := M_h,
\end{align*}
where 
$\mathscr{P}_1(K)$ is the space of linear functions on $K \in\mathscr{T}_h$.
\par
We now introduce notations to be used in the paper.
We denote by $(\cdot, \cdot)$ the $L^2(\Omega)$-inner product, that is,
\[
(f, g) := \int_{\Omega} f(x) g(x) \, d{\bf x}, \quad (f, g \in L^2(\Omega)),
\]
and use the same notation as the inner products in $L^2(\Omega)^2$ and $L^2(\Omega)^{2\times 2}$, the vector- and matrix-valued $L^2(\Omega)$ function spaces.
Let $a$, $b$ and $\mathscr{C}_h$ be bilinear forms defined by
\begin{align*}
a({\bf u}, {\bf v}) & := \frac{2}{Re} \bigl( D({\bf u}), D({\bf v}) \bigr), \\
b({\bf v}, q) & := -(\nabla\cdot {\bf v}, q), \\
\mathscr{C}_h(p, q) & := \delta_0 \sum_{K\in\mathscr{T}_h} h_K^2 \int_K \nabla p \cdot \nabla q \, d{\bf x},
\end{align*}
where
$\mathscr{C}_h(p, q)$ is Brezzi--Pitk\"aranta's pressure-stabilization term\cite{BrePit-1984},
$\delta_0 > 0$ is a stabilization parameter,
and $h_K := {\rm diam} (K)$ is the diameter of $K \in \mathscr{T}_h$.
\par
We have employed the stabilized LG method{\it\cite{Notsu-Tabata}}; 
find $\{ ({\bf u}^n_h,p^n_h) \in V_h({\bf u}_\ast^n) \times Q_h; \ n=1, \ldots, N_T \}$ such that
for $n=1, \ldots, N_T$,
\begin{subequations}\label{scheme}
\begin{align}
\Bigl( \tfrac{{\bf u}_h^n - {\bf u}_h^{n-1} \circ X_1({\bf u}_h^{n-1},\Delta t)}{\Delta t}, {\bf v}_h \Bigr) \  \notag\\
+ a( {\bf u}_h^n, {\bf v}_h) + b( {\bf v}_h, p_h^n) =&\, 0, \ \ \forall {\bf v}_h \in V_h, \\
b({\bf u}_h^n, q_h) - \mathscr{C}_h(p_h^n, q_h) = &\, 0, \ \ \forall q_h \in Q_h, 
\end{align}
\end{subequations}
where
$X_1({\bf w}_h, \Delta t): \Omega \to \mathbb{R}^2$ is a mapping defined by
\[
X_1({\bf w}_h, \Delta t)({\bf x}) := {\bf x} - {\bf w}_h({\bf x}) \Delta t
\]
for a velocity ${\bf w}_h: \Omega \to \mathbb{R}^2$,
which is the so-called {\em upwind point} of ${\bf x}$ with respect to ${\bf w}_h({\bf x})$,
and the symbol~$\circ$ is the composition of functions, that is,
\[
{\bf v}\circ X_1({\bf w},\Delta t)({\bf x}) := {\bf v} ( X_1({\bf w},\Delta t)({\bf x}))
\]
for velocities ${\bf v}$ and ${\bf w}$.
The parameters used in the simulation are listed in Table~\ref{tab:params}.
Notably, the convergence property with error estimates of the scheme in an ideal setting have been proved in~{\it\cite{Notsu-Tabata}}, and the scheme has been employed to understand tornado-type flows with vortices governed by the three-dimensional Navier--Stokes equations~{\it\cite{HNY-2016}}.
\begin{table}[]
\centering
\begin{tabular}{ccc}\hline
Symbol& Parameter & Value \\ \hline
$N_p$ & \#nodes in flow Sim. & 2,734 \\
$N_e$ & \#elements in flow Sim. & 5,278 \\
$N_r$ & \#nodes of output/observation & 231 \\
$\Delta t$ & Time increment in flow Sim. & 0.2 \\
$\delta_0$ & Stabilization parameter & 0.1 \\
$\tau$ & Transient time in flow Sim.& 5 \\
$\alpha$ & Tilt parameter & 0.5 \\
$T_i$ & \#inputs & 2,000 \\ 
$T_{ii}$ & \#initial inputs & 100 \\
$T_w$ & Washout time in reservoir& 100 \\
$T_t$ & Training time in reservoir& 1,200 \\
$T_e$ & Evaluation time in reservoir& 700 \\
$T_r$ & Total time in reservoir & $T_i+T_{ii}$ \\
$T_f$ & Total time in flow Sim. & 10,500 \\ \hline
\end{tabular}
\caption{Values of parameters}
\label{tab:params}
\end{table}

\section*{Amplification factor of the perturbation
}
We have defined $\lambda$ in the main text as an indicator of stability of the solution to Navier--Stokes problem.
In this section, we introduce the complete procedure to obtain the value of~$\lambda$.
\par
Let $({\bf u}_\dagger, p_\dagger): \Omega\times [0, T_f] \to \mathbb{R}^2\times\mathbb{R}$ be the solution to Navier--Stokes problem.
Considering the linearization of the Navier--Stokes equations near $({\bf u}_\dagger, p_\dagger)$, we derive a problem to find the perturbation $(\tilde{\bf{u}},\tilde{p}): \Omega \times (0,T_f)\to\mathbb{R}^2\times \mathbb{R}$ such that
\footnotesize
\begin{align*}
\frac{\partial\tilde{{\bf u}}}{\partial t} + ({\bf u}_\dagger \cdot \nabla)\tilde{{\bf u}} + (\tilde{{\bf u}} \cdot \nabla){\bf u}_\dagger \\
- \nabla \cdot \sigma(\tilde{{\bf u}}, \tilde{p}) & = {\bf 0} && \!\! \mbox{in}\ \Omega \times (0,T_f),\\
\nabla \cdot \tilde{{\bf u}} & = 0 && \!\! \mbox{in} \ \Omega \times (0,T_f), \\
[\sigma(\tilde{{\bf u}},\tilde{p}){\bf n}] \times {\bf n}^\perp = 0, \ \tilde{{\bf u}} \cdot {\bf n} & = 0 && \!\!\! \mbox{on}\ \Gamma_{2} \times (0,T_f), \\
\sigma(\tilde{{\bf u}},\tilde{p}){\bf n} & = {\bf 0} && \!\!\! \mbox{on}\ \Gamma_{3} \times (0,T_f), \\
\tilde{{\bf u}} & = {\bf 0} && \!\!\! \mbox{on}\ (\Gamma_{1}\cup\Gamma_{4}) \times (0,T_f), \\
\tilde{{\bf u}} & = \tilde{{\bf u}}^0 && \!\! \mbox{in}\ \Omega,\ \mbox{at}\ t=0,
\end{align*}
\normalsize
where $\tilde{{\bf u}}^0$ is a given initial perturbation approximated by $\tilde{{\bf u}}^0_h \in V_h$ and defined by
\begin{align}
\tilde{{\bf u}}^0_h (P) & := 
\left\{
\begin{aligned}
& {\bf 0} && (P \neq P_\ast), \\
& (0, 0.01)^T && (P = P_\ast),
\end{aligned}
\right.
\label{def:initial_perturbation}
\end{align}
for  nodes $P$ and $P_\ast := (-7.5, 0)$
\par
Let $\{ ({\bf u}_{h\dagger}^n, p_{h\dagger}^n) \in V_h({\bf u}_\ast^n) \times Q_h; n=1, \ldots, N_T\}$ be the solution to scheme~\eqref{scheme}.
To obtain an approximate value of $\lambda$, we use a numerical scheme;
find $\{ (\tilde{{\bf u}}^n_h, \tilde{p}^n_h) \in V_h \times Q_h; \ n=1,...,N_T \}$ such that, for $n=1, \ldots, N_T$,
\begin{subequations}\label{scheme1}
\begin{align}
\Bigl( \tfrac{\tilde{{\bf u}}_h^n - \tilde{{\bf u}}_h^{n-1} \circ X_1({\bf u}_{h\dagger}^n,\Delta t)}{\Delta t}, {\bf v}_h \Bigr) 
+ & \bigl( (\tilde{{\bf u}}_h^{n-1} \cdot \nabla){\bf u}_{h\dagger}^n, {\bf v}_h \bigr) \notag\\
+ a( \tilde{{\bf u}}_h^n, {\bf v}_h) + b( {\bf v}_h, \tilde{p}_h^n) = & \, 0, \ \ \forall {\bf v}_h \in V_h, \\
b(\tilde{{\bf u}}_h^n, q_h) - \mathscr{C}_h(\tilde{p}_h^n, q_h) = & \, 0, \ \ \forall q_h \in Q_h, 
\end{align}
\end{subequations}
with the initial perturbation $\tilde{{\bf u}}_h^0$ defined by definition~\eqref{def:initial_perturbation}.
Notably, scheme~\eqref{scheme1} has convergence property with error estimates proved in~{\it\cite{NotTab-2015}}.
\par
From numerical experiments, the critical Reynolds number~$Re_c$ of the onset of (wake) instability of flow with the time-independent function ${\bf z}_\ast=(1,0)^T$ in our Navier--Stokes problem is in a range $Re_c \in [40, 50]$(cf. e.g.,{\it\cite{Takemot-Mizushima}} and the references therein).

\section*{Ridge regression}
We use ridge regression{\it\cite{Hastie-Tibs}} following $L^2$ regularization in the process of learning. The detailed steps are as follows. Ridge regression is used to minimize the residual sum of squares between system and target outputs by penalizing the size of the weights in the training phase. When the training phase is defined as time domain $200 \le k \le 1399$ and the optimal weights are $w_{out}^i$ , the problem to be solved is ~$\beta^i = \rm argmin_{w}\{\sum_{k=200}^{1399}(y_k-\hat y_k)^2 + \lambda_R \sum_{i=0}^{N_r}(w_{out}^i)^2\}$, where $\lambda_R \in \mathbb{R}_{\ge0}$ is the ridge parameter. By collecting a training data set of $1200$ timesteps, we obtain a $1200 \times N_r$ matrix ${\bf X}$, where $N_r$ is the number of nodes in the reservoir. Furthermore, the target output corresponding to $1200$ timesteps is determined by ${\bf y}=[y_{200},...,y_{1399}]^T$. Then we can obtain the optimal weights,  ${\bf W_{out}}=[w_{out}^0,...,w_{out}^{N_r-1}]^T$, by:
\begin{align*}
{\bf W_{out}}=({\bf X}^T {\bf X} +\lambda_R {\bf I})^{-1}{\bf X}^T{\bf y}, 
\end{align*}
We must determine the appropriate $\lambda_R$, for which the degrees of freedom ($df$) is crucial. The value of $df$ is determined by
\begin{align*}
df = \sum_{i=1}^{N_r} \frac{\lambda_i}{\lambda_i + \lambda_R},
\end{align*}
where $\{\lambda_i\}$ are the eigenvalues of ${\bf X}^T {\bf X}$. Because $df$ is a one-to-one correspondence with $\lambda_R$
and its maximum value is $N_r$, we can obtain unique $\lambda_R$ by increasing $df$ from $1$ to $N_r$ by one, and corresponding the ${\bf W_{out}}$ can also be obtained. The residual sum of squares $RSS = \sum_{k=200}^{1399}(y_k-\hat y_k)^2 $
can be calculated by using  ${\bf W_{out}}$, and we can obtain Akaike's Information Criterion($AIC$) as follows:
\begin{align*}
AIC \coloneqq M \log (RSS) + 2df,
\end{align*}
where $M$ is a training size of $1200$ in our research. By calculating $AIC$ according to each $df$, the optimal  $\lambda_R$ and ${\bf W_{out}}$ that minimize $AIC$ can be obtained.
 
\section*{Comparisons with ESN}
In the main text, to characterize the computational capability of our reservoir, we compared its performance in the NARMA tasks and its memory capacities with those of a conventional ESN{\it\cite{Jaeger-Haas,Jaeger1,Jaeger2}}.
We explain the settings of the ESN used for the comparisons in detail.

The ESN is a recurrent neural network comprising internal computational nodes (containing $N_{\rm ESN}$ nodes), input nodes, and output nodes. 
The activation of the $i$-th internal node at timestep $k$ is expressed as $x^{i}_{k}$.
The weights $w^{ij}$ for the internal network connect the $i$-th node to the $j$-th node and are randomly determined from the range $[-1.0, 1.0]$, and the spectral radius of the weights is adjusted and fixed to $0.9$ throughout the experiments.
The input weights $w^{i}_{in}$ connect the input node to the $i$-th internal node and are randomly determined from the range $[-\sigma, \sigma]$, where $\sigma$ is a scaling parameter controlled for each task, which we explain in this section later.
The internal nodes with one bias are connected to the output unit through readout weights $w^{i}_{out}$, where $x^{0}_{k} = 1$ and $w^{0}_{out}$ are assigned as the bias term.
Taking the activation function as $f(x) = \tanh(x)$, the time evolution of the ESN is expressed as follows:
\begin{align}
x^{i}_{k} &= f(\sum^{N_{\rm ESN}}_{j=1}w^{ij}x^{j}_{k-1} + w^{i}_{in}z_{k}), \\
y_{k} &= \sum^{N_{\rm ESN}}_{i=0}w^{i}_{out}x^{i}_{k}.
\end{align}
Learning is performed by adjusting the linear readout weights $w^{i}_{out}$ based on the same procedure using the ridge regression explained in the above section.
To conduct a fair comparison of the task performance, the Input/Output (I/O) settings and the evaluation procedures were also kept the same.
Because we used a larger number of computational nodes for ESN than our Karman vortex reservoir in some conditions, the lengths of the washout, training, and evaluation phases were set accordingly longer to avoid over-fitting and to conduct safe comparisons.  
The detailed experimental conditions for each task are as follows.

For the NARMA task, we first prepared 20 different ESNs for each setting of $N_{\rm ESN}$ (Here, we performed the experiments with $N_{\rm ESN} = 10,$ and $600$. Only the performance of the relevant conditions of $N_{\rm ESN}$ are shown for comparisons in Fig. 3).
The lengths of the washout, training, and evaluation phases are set as 2000, 5000, and 5000 timesteps, respectively.
The scaling parameter of the input weights $\sigma$ is fixed to 0.01.
For each ESN, we ran ten different trials starting from different initial states and tested the emulation tasks of all the NARMA systems using a multitasking scheme for each trial.
After performing all the trials of the NARMA tasks for each ESN having the computational node $N_{\rm ESN}$, we calculated the averaged {\it NMSE} for each NARMA task using the ten different trials, and then the {\it NMSE} value is further averaged using 20 different ESNs for each setting of $N_{\rm ESN}$.
These averaged {\it NMSE}s were used for comparison.   

Similar to the NARMA task, to evaluate the memory capacities, 20 different ESNs (for each condition with $N_{\rm ESN} = 5, 10, 50,$ and $200$) were prepared.
The lengths of the washout, training, and evaluation phases are set as 2000, 5000, and 5000 timesteps, respectively.
The averaged nonlinear memory capacities are calculated in a similar manner explained for the NARMA task.
In {\it\cite{Dambre-Vers}}, it is reported that ESN can only demonstrate nonlinear memory capacities with odd degrees because the hyperbolic tangent is an odd function and, according to the increase of the input scaling, the ESN becomes increasingly nonlinear.
Thus, the nonlinear memory capacities for even degrees could not be obtained in the analysis and presented in Fig. 3.
Based on the insight, we varied the scaling parameter of the input weights by $\sigma=0.01, 1.0,$ and $5.0$ and used conditions that demonstrate the highest capacities among them on the basis of the same number of $N_{\rm ESN}$ for each comparison in Fig. 3.
We also observed that according to the increase in the scaling parameter of the input weights, the higher nonlinear memory capacities were obtained on the basis of the same number of $N_{\rm ESN}$.
Namely, for the linear memory capacity (nonlinear degree 1), the results for the $\sigma=0.01$ were the highest and used for the comparisons.
Subsequently, for the nonlinear memory capacity of degree three and five, the results for the $\sigma=1.0$ and $5.0$ showed the highest, respectively, and used for comparisons in Fig. 3.

For the prediction task for the fluid variables, similar to the aforementioned tasks, 20 different ESNs (for each condition with
$N_{\rm ESN}= 5,$ and $200$) were prepared and the same I/O target data set with our vortex-reservoir system are used for the analysis (containing eight different trials).
The experiments are conducted for each variable and for each Reynolds number by following the same procedure as our experiments conducted on the vortex-reservoir system and the averaged {\it NMSE}s are obtained in a manner similar to that explained for the NARMA task for each condition.

\renewcommand{\refname}{References}